# Ferroelastic twin reorientation mechanisms in shape memory alloys elucidated with 3D X-ray microscopy

A.N. Bucsek[1†], D.C. Pagan[2], L. Casalena[3‡], Y. Chumlyakov[4], M.J. Mills[3], A.P. Stebner[1*]

[1]Mechanical Engineering, Colorado School of Mines, 1610 Illinois St., Golden, CO 80401.

[2]Cornell High Energy Synchrotron Source, 161 Wilson Laboratory, Ithaca, NY 14853.

[3]Materials Science Engineering, The Ohio State University, 2041 N. College Rd., Columbus, OH 43210.

[4]Department of Physics of Metals, Tomsk State University, Lenin Ave. 36, Tomsk, Russia 634050.

*Corresponding author: astebner@mines.edu.

[†]Present address: Aerospace Engineering and Mechanics, University of Minnesota, 110 Union St. SE, Minneapolis, MN 55455.

[‡]Present address: Thermo Fisher Scientific, 5350 NE Dawson Creek Dr., Hillsboro, OR 97142.

**Abstract:** Three-dimensional (3D) X-ray diffraction methods were used to analyze the evolution of the load-induced rearrangement of monoclinic twin microstructures through bulk nickel-titanium specimens in 3D and across three orders of magnitude in length scales: changing lattice plane spacing and orientation at the nanoscale, growth and nucleation of martensite twin variants at the microscale, and localization of plastic strain into deformation bands at the macroscale. Portions of the strain localization bands were reconstructed in situ and in 3D. Analyses of the data elucidate the sequence of twin rearrangement mechanisms that occur within the propagating strain localization bands, connect these mechanisms to the texture evolution, and reveal the effects of geometrically necessary lattice curvature across the band interfaces. The numerous similarities between shear bands and localized deformation bands in twin reorientation are also discussed. These findings will guide future researchers in employing twin rearrangement in novel multiferroic technologies, as well as demonstrate the strength of these types of 3D, multiscale, in situ experiments for improving our understanding of complicated material behaviors and providing opportunities to accelerate our abilities to model them.

**Highlights:**
- Strain localization initiates via detwinning and twin nucleation
- Strain localization saturates via detwinning only

- Strain localization leads to geometrically necessary lattice curvature
- Lattice curvature influences and may even enable complete twin rearrangement
- 3DXRD advancements are used to connect micro- and macro-scale mechanics

**Keywords:** shape-memory materials, ferroics, stimuli-responsive materials, structure-property relationships, characterization tools

**1. Introduction**

The functional properties of ferroic materials have driven them to the forefront of advanced materials research for innovating technologies in energy harvesting, biomedical applications, microelectromechanical systems (MEMS), ferroelectric random access memory (FRAM or FeRAM), and many others (Bowen et al., 2014; Kohl, 2004; Muralt, 2000; Trolier-McKinstry and Muralt, 2004). Ferroic materials respond to external stimuli by undergoing spontaneous changes in macroscopic behavior: ferroelastics, ferroelectrics, and ferromagnetics undergo spontaneous changes in mechanical, electrical, and magnetic ordering, respectively. In particular, ferroelastic materials can accommodate large mechanical loads and deformations without damage through reversible rearrangement of their crystal structure via phase transformations and/or twin reorientation. Ferroelasticity is the fundamental behavior underlying the sizeable reversible strains exhibited by shape memory alloys (SMAs) (Bhattacharya et al., 2004; Bhattacharya and James, 2005). Most ferroelectrics (Cruz et al., 2007; Feigl et al., 2014; Li et al., 2016; Mascarenhas et al., 2017; Phillips et al., 2015) and many ferromagnetics (Heczko et al., 2000; Lahtinen et al., 2012; Liu et al., 2016; Omori et al., 2011) are also ferroelastic, making such materials "multiferroic." Multiferroic coupling results in an ability to control a material's electrical and/or magnetic properties with mechanical deformation, and vice versa. For example, when crystallographic twin domains rearrange ferroelastically in ferroelastic-ferroelectric materials, including $BaTiO_3$, $Pb(Zr,Ti)O_3$, and $BiFeO_3$, the macroscopic electrical properties and strain of the material change simultaneously. These reversible changes can be induced by either switching an electrical field or cycling a mechanical load. Similarly, when magnetic fields or mechanical loads are cyclically applied to ferroelastic-ferromagnetic materials, such as NiMnGa, FeGaB, and FePd, commonly known as ferromagnetic SMAs, the macroscopic magnetic properties and strain of the material change simultaneously. Hence, ferroelastic twin rearrangement is the enabling mechanism for a wide variety of novel functional material technologies in energy conversion, information storage media, and sensor and actuator applications (Ramesh and Spaldin, 2007; Scott, 2007; Setter et al., 2006).

Crystallographic *twins* consist of two crystals of the same structure that meet at shared planes called the *twin planes* (Wechsler et al., 1953). If one crystal structure "looked" across twin plane, the crystal structure on the other side of the plane would be its mirrored reflection, or twin (Wechsler et al., 1953; Bilby and Crocker, 1965). Twins can also be related through a simple shear deformation of one side of the twin with respect to the other. The two crystal structures that make up a twin are different crystallographic orientations called *variants*. Variants are symmetry-related, but they are rotationally unique—it is not possible to map one variant to another via pure rotation. In the absence of an applied load, twins will form *self-accommodation twins* that fit together to accommodate, or preserve the macroscopic volume and shape. In the presence of an applied load, the applied load will bias the twins that form, and the twins will (re)arrange to maximize the order parameter (strain, polarization, or magnetization) in the loading direction. This twin rearrangement is commonly called twin *reorientation*, or martensite reorientation in SMAs since the twinned phase is the martensite phase. Two special subclasses of twinning operations have been proposed to be the principal mechanisms of twin reorientation: *twin nucleation* - a twinning mechanism in which a new type of twin nucleates from a preexisting twin of a different type and *detwinning* - a twinning mechanism in which the volume of one variant of a twin grows as the volume of the other variant diminishes, but the type of twin does not change (Müllner and King, 2010). If a twinned material is sufficiently deformed, then all of the twins will eventually reorient to the *favorable variant*. In mechanical loading of ferroelastic materials, the favorable variant is the variant that maximizes the mechanical work, or in the case of uniaxial tension, maximizes the strain in the loading direction. When the microstructure has completely reoriented to the favorable variant, it is said to be *fully reoriented*. Additional terms used in this work are:

- *Domain*: a continuous volume consisting of a single crystallographic orientation, or variant
- *Grain*: a volume that is surrounded by grain boundaries (not to be confused with twin planes). In a twinned material, a grain may contain many twins within it. When a ferroelastic material has fully reoriented to the favorable variant, a grain will typically become a continuous domain of the crystallographic orientation associated with the favorable variant.
- *Localized deformation band (LDB)*: a macroscopic band-like structure of significant local deformation relative to the rest of the specimen

Previous studies of twin reorientation in SMAs fall into two main categories: nano- and micro-scale surface or near-surface studies of microstructure (Liu, 2001; Liu et al., 2000, 1999, 1998; Muntifering et al., 2016; Tadayyon et al., 2017) and in situ bulk measurements averaged over hundreds to millions of grains (Dilibal, 2013; Laplanche et al., 2017; A. Stebner et al., 2013; Stebner et al., 2011). From nano- and micro-scale surface experiments, we have evidence of individual reorientation mechanisms by way of small "snapshots" of the microstructure and have a general idea of where they occur relative to the macroscopic stress-strain response. Using ex situ transmission electron microscopy (TEM), researchers have shown that twins can inelastically reorient even before the onset of the stress plateau (Liu et al., 2000; Tadayyon et al., 2017), that the nucleation of (100) compound twins from Type-II twins occurs early in the deformation process (Liu et al., 2000), and that the plateau corresponds to large-scale reorientation and detwinning (Liu, 2001; Liu et al., 2000; Tadayyon et al., 2017). From in situ bulk experiments, we understand the macroscopic stress-strain response in terms of propagating LDBs and averaged texture changes. Using in situ digital image correlation (DIC) measurements, researchers have shown that there are LDBs that coincide with the stress plateau (Dilibal, 2013; Laplanche et al., 2017). Using neutron diffraction studies, researchers have shown how the twinned martensite texture averaged over tens of millions of grains evolves with loading and related that texture evolution to the roles of different elastic, recoverable reorientation twinning, unrecoverable deformation twinning, and plastic deformation not due to twinning (A. Stebner et al., 2013; Stebner et al., 2011).

In spite of these recent advancements in micromechanical understanding of reorientation processes, the gap between these two types of measurements, ex situ surface micrographs and in situ polycrystalline averages, has left many open questions in our understanding of load-induced twin rearrangement. For example, despite in situ evidence of twin nucleation (Muntifering et al., 2016), many micromechanical models of load-induced reorientation stress-strain curves do not include twin nucleation (Dilibal, 2013; Liu, 2001; Tadayyon et al., 2017). There is also little direct evidence of how twin nucleation and detwinning interact within a LDB, though it has been insightfully hypothesized (Laplanche et al., 2017). Three-dimensional (3D) characterization of the internal structure of twin reorientation LDBs is also missing—only 2D surface observations exist, and because the crystals are monoclinic, it is not possible to infer out-of-plane structures. There is almost no evidence regarding if and how elastic stress heterogeneities may emerge and interact with the reorientation process because previous measurements are either ex situ or lack sufficient spatial resolution. Finally, tracking the volumes and numbers of each variant or twin system throughout a specimen during deformation remains an open challenge. Such unprecedented observations together with quantified statistics of reorientation would be extremely valuable for both verifying the ubiquitousness of observed surface events, as well as informing and verifying 3D models of reorientation processes.

In this study, we bridge the gap between ex situ micrographs and in situ bulk diffraction measurements using a suite of 3D X-ray diffraction (3DXRD) methods (Poulsen, 2004) to measure sequential snapshots of ferroelastic twin reorientation within a bulk material volume in situ, in 3D, and across six orders of magnitude in spatial length scales: changing lattice plane spacing and orientation at the nanoscale, growth and nucleation of martensite twin variants at the microscale, and localization of plastic strain into deformation bands at the macroscale. The two X-ray diffraction-based methods are called far-field and near-field high-energy diffraction microscopy (ff- and nf-HEDM). Far field-HEDM can be used to measure the 3D orientation, position, volume, and elastic strain of individual grains, while nf-HEDM can be used to spatially map grain morphologies. HEDM and similar 3D X-ray methods including X-ray diffraction contrast tomography, topotomography and phase contrast tomography have been successfully applied to studying shear banding and deformation banding (Gueninchault et al., 2016; Pagan et al., 2018; Proudhon et al., 2018), and as twinning (Abdolvand et al., 2015; Bucsek et al., 2018; Lind et al., 2014; Nervo et al., 2016; Paranjape et al., 2018; Stein et al., 2014; Viganò et al., 2016). This study lies at the intersection of these two areas. The ensuing text presents new micromechanical understanding that could not be previously gained from more traditional in situ powder diffraction studies or ex situ microscopy methods.

Specifically, we report on three martensitic nickel-titanium (NiTi) SMA samples, where the B19′ monoclinic martensite phase has 12 variants. At high temperatures during manufacturing, the samples initially consisted of B2 cubic austenite grains (ranging from 5 to 15 in different specimens) that were misaligned by 1°–18° (i.e., near-single crystals). Upon cooling to room temperature, each parent grain transformed to many (thousands or millions) martensite self-accommodation twins. Because the microstructure consisted of a limited number of similarly oriented (in the B2 austenite sense) grains, it was possible to uniquely identify each of the 12 martensite variants in the X-ray diffraction data and to track their behaviors throughout deformations (see **Appendix A.3.2** for details). The initiation and propagation of macroscopic LDBs were measured with DIC, and subsequently correlated with the variant interactions within the microstructures to study twin nucleation and detwinning interaction mechanisms in situ. Reoriented martensite microstructures were spatially resolved using nf-HEDM, revealing the 3D internal structures of LDBs. Finally, corresponding lattice rotations and elastic strain heterogeneities were quantified from the ff-HEDM data, revealing an elastic lattice misorientation mechanism associated with the propagating LDBs.

## 2. Materials and methods
### 2.1 Experiment overview

Three B19' monoclinic martensite NiTi samples were prepared and cut into test specimen with 1×1 mm$^2$ square-shaped cross-sections (**Fig. 1**). Each sample is polycrystalline, and contains between 5 and 15 similarly oriented grains within the 1×1×1 mm$^3$ central regions of the gage sections that were illuminated by X-rays during ff-HEDM measurements. (Recall that 'grain' defines the domains of the samples that were of the same B2 orientation prior to transforming to martensite, as defined in **Section 1**. Here, we classify boundary misorientation values ≥ 2° as grain boundaries.) Samples 1 and 2 had an 8 mm long gage section, and sample 3 had a 1 mm long gage section (**Table 1**). The specimen with the shorter gage section was included to observe any differences caused by a more concentrated load.

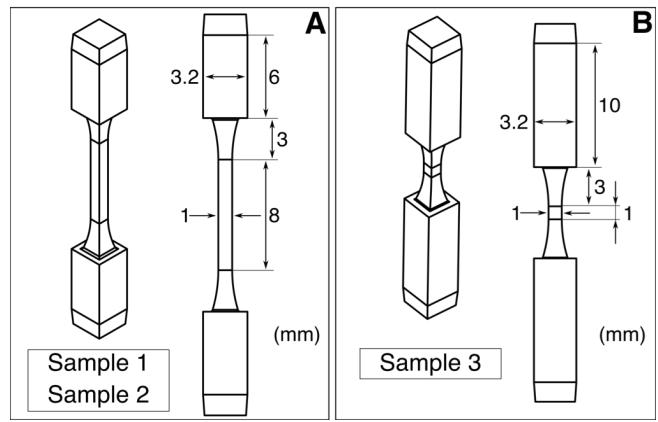

**Fig. 1.** Specimen geometry for samples 1 and 2 (A) and for sample 3 (B).

**Table 1.** Specimen dimensions and load steps at which different measurements occurred for samples 1, 2, and 3.

| | Sample 1 | Sample 2 | Sample 3 |
|---|---|---|---|
| Gage section dimensions | 1×1×8 mm$^3$ | 1×1×8 mm$^3$ | 1×1×1 mm$^3$ |
| ff-HEDM measurement step #s | 0–7 | 0–5 | 0–4 |
| Volume illuminated by X-rays during ff-HEDM measurements | 1×1×1 mm$^3$ | 1×1×1 mm$^3$ | 1×1×1 mm$^3$ |
| nf-HEDM measurement step # | 6 | 5 | 4 |
| Volume illuminated by X-rays during nf-HEDM measurements | 1×1×1 mm$^3$ | 1×1×0.5 mm$^3$ | 1×1×0.5 mm$^3$ |
| Lattice rotation vs. elastic strain calculation step #s | 0–6 | 0–5 | N/A |

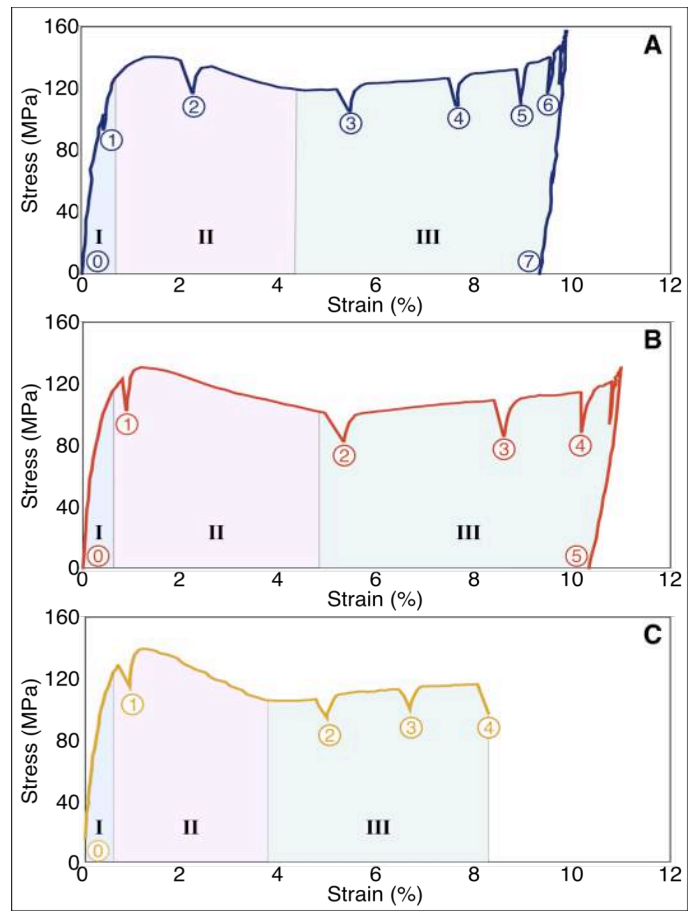

**Fig. 2.** Stress-strain responses for sample 1 (A), sample 2 (B), and sample 3 (C). The stress-strain response show three general regions of behavior labeled I, II, and III. The stress relaxation "spikes" occurred as a result of fixing the load frame crosshead during X-ray measurements. Note: sample 3 did not fracture—DIC images were not recorded as the sample was unloaded from load step 4.

To stress-induce ferroelastic twin reorientation, each of the three samples was loaded in tension to 8–11% macroscopic strain and then unloaded using crosshead displacement control (see **Fig. 2**). The 8–11% strain limits were chosen to limit the focus of this study to ferroelastic reorientation mechanisms, as larger deformations typically activate significant amounts of plasticity and non-ferroelastic twinning mechanisms (A. Stebner et al., 2013). Digital images of the sample surfaces were recorded during deformation to quantify surface strains on one face of the gage section using DIC methods.

As further described in **Section 2.4**, nf- and ff-HEDM measurements were taken at the load steps indicated in **Table 1**, where the load step numbers to the labels shown in **Fig. 2**. The ff-HEDM measurements were used to measure the texture, calculate the volume of each of the 12 variants, and measure the lattice rotation vs. corresponding elastic strain at each load step. The nf-HEDM measurements were used to spatially resolve 3D orientation maps of the fully detwinned martensite structures.

**2.2 Materials and sample preparation**

Three samples were electrical-discharge-machined (EDM) from a 40 mm diameter B19′ monoclinic martensite NiTi ingot. The ingot was grown by an advanced Bridgman crystal growth technique consisting of remelting cast and extruded $Ni_{50.1}Ti_{49.9}$ rods into a graphite crucible under an inert helium gas atmosphere. This crystal growth produced large, slightly misoriented B2 cubic austenite grains that then transformed to twinned B19′ monoclinic martensite structures upon cooling to room temperature. High-angle annular dark-field (HAADF) scanning transmission electron microscopy (STEM) micrographs taken using an FEI Probe Corrected Titan3 at 300 kV show examples of the initial monoclinic twin structures within the samples (**Fig. 3**). **Fig. 3** confirms that the microstructure initially contains hierarchical twins consisting of a variety of twin orientations, twin types, and twin length scales.

Two specimen geometries were used for the three samples. The specimen geometry for the samples designated 1 and 2 has a 1×1×8 $mm^3$ gage section (**Fig. 1A**). This specimen geometry is designed specifically for the second generation of the Rotational and Axial Motion System (RAMS2) load frame (Turner et al., 2016). The specimen geometry for the sample designated 3 has a 1×1×1 $mm^3$ gage section (**Fig. 1B**). This specimen geometry was modified from that of (Turner et al., 2016) to produce a more concentrated load in the illuminated volume,

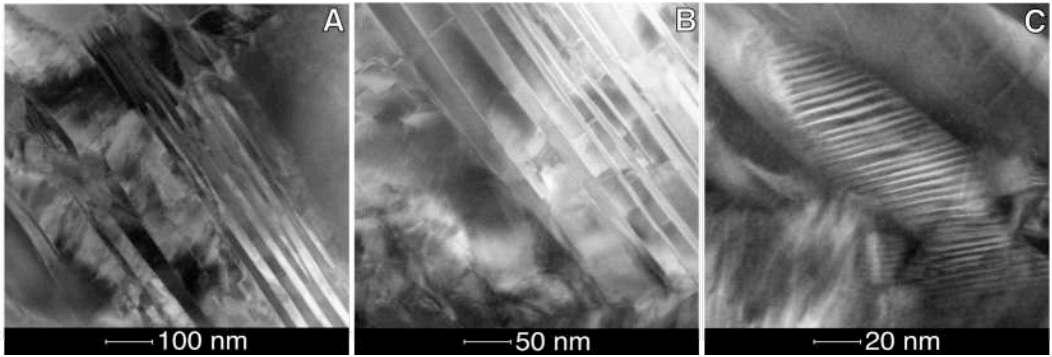
**Fig. 3.** Three different HAADF STEM micrographs of a section taken from the grip of sample 1 showing that variations of the initial martensite twin morphologies span multiple length scales, where (A) has a 100 nm scale bar, (B) has a 50 nm scale bar, and (C) has a 20 nm scale bar.

with the goal of ensuring that the initiation of the LDB would remain within the X-ray diffraction volume. Experiment time and hardware limitations made in situ 3DXRD measurements of entire 8 mm long gage sections impractical.

### 2.3 Mechanical loading

The three samples were loaded in tension to stress-induce ferroelastic twin reorientation (**Fig. 2**). The mechanical loading was performed using the RAMS2 load frame, a load frame capable of applying tension and compression loads to samples while the entire load train rotates continuously about the central axis of the samples, constructed specifically for HEDM measurements (Schuren et al., 2015; Shade et al., 2015). Using RAMS2, tensile loads were quasi-statically applied to the samples using displacement-controlled movements of the crosshead at a rate of 0.001 mm/s. At multiple points in the loading, the loading was paused and nf- and/or ff-HEDM measurements were taken (see **Table 1**). The load train of the RAMS2 load frame is mounted on air bearings within the frame, which allows for continuous $\omega$ rotation ($\omega$ is defined in **Fig. 4** in the next section) of the sample without shadowing from load frame columns. During HEDM measurements, the crosshead was fixed in displacement, and the sample was rotated 360º in $\omega$.

To measure the macroscopic strain of specimens during loading, image series of the as-EDM-prepared specimen surfaces (i.e., no speckle pattern was applied) were recorded using a FLIR (Point Grey) Grasshopper GS3-U3-50S5M-C camera, which has a Sony ICX625 2448 × 2048 (2/3") CCD with 3.45 μm pixel size sensor. A Standard & Precision Optics TCL0.8X-110-HR lens with a 110 mm working distance, 14.9 μm resolution, and 0.0255 numeric aperture was used. The displacement fields and strains were calculated from these images using Ncorr digital image correlation (DIC) software (Blaber et al., 2015). The strain for the stress-strain curves in **Fig. 2** show the average strain from only the area within the 1×1×1 mm$^3$ illuminated volume.

### 2.4 In situ X-ray measurements

In situ nf- and ff-HEDM measurements were performed at the F2 beamline of the Cornell High Energy Synchrotron Source (CHESS). A typical HEDM setup is shown in **Fig. 4**. The far-field detector is located ~1 m away from the sample and is used to collect ff-HEDM data. The near-field detector is located 5–15 mm from the sample and is used to collect nf-HEDM data. **Appendices A** and **B** describe the specific HEDM setups and techniques used in this work. In HEDM measurements, a monochromatic X-ray beam and a 360° sample rotation $\omega$ about the loading axis $\mathbf{y}_L$ are typically used (Lauridsen et al., 2001; Lienert et al., 2011; Poulsen, 2012, 2004; Poulsen et al., 2001). Samples are designed such that only a few (one to a few thousand) crystals satisfy a Bragg condition at each $\omega$ rotation increment (typically 0.1–0.5°), and the Bragg reflections of one crystal are distinguishable from the others. Then, a diffraction pattern is collected at each $\omega$ rotation increment. Instead of acquiring orientation-averaged information as is done in powder diffraction measurements, this extra "dimension" of analyzing each Bragg reflection of each grain allows for the unique analysis of each individual grain. When the detector is in the near-field configuration, the grain topology dominates the diffraction pattern and quantities like relative grain shape, size, location, and orientation can be resolved (much like a 3D electron backscatter diffraction measurement except non-destructively). When the detector is in the far-field configuration, the reciprocal-space distribution of lattice strains and orientations dominates the diffraction pattern and quantities like grain-specific lattice strains, orientation distributions, positions, and volumes can be resolved. The nf-HEDM orientation and spatial resolution are typically quoted as 0.1° and 2 μm, respectively (Li et al., 2012), and the ff-HEDM strain resolution has been measured as $\pm 1 \times 10^{-4}$ (Pagan et al., 2018).

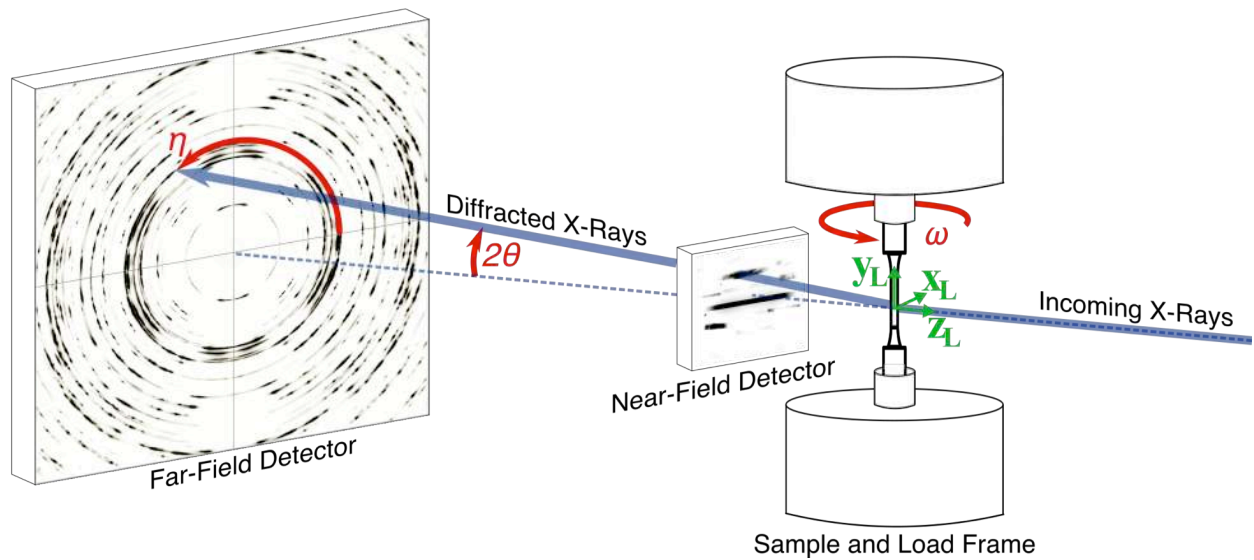

**Fig. 4.** Schematic of the HEDM experimental setup including the laboratory coordinate system, $\mathbf{x}_L$, $\mathbf{y}_L$, $\mathbf{z}_L$. The diffraction angles, $\omega$, $2\theta$, and $\eta$, at which a Bragg reflection is recorded define the orientation and interplanar spacing of the diffracting plane.

To date, most HEDM analyses have been performed on high symmetry materials with large (20–200 μm), uniform grain sizes. Twinned materials can significantly increase the difficulty of HEDM data analyses due to the small domain sizes that often accompany twins, though several researchers have made great strides toward using HEDM to study the Schmid ranking of twinning systems in zirconium polycrystals (Abdolvand et al., 2015; Lind et al., 2014), NiTi single crystals (Bucsek et al., 2018; Paranjape et al., 2018), and heterogeneous grain rotation in actuated NiTi (Bucsek et al., n.d.). Low symmetries can also add significant challenges to HEDM analysis (Bucsek et al., 2018). The materials in this study are both initially highly twinned and low symmetry (monoclinic). In light of this, several novel approaches to HEDM data analysis and visualization were developed for this study. We refer interested readers to **Appendix A.3.2** for details on the variant volume histogram calculations, **Appendix A.3.3** for details on the subgrain-scale lattice rotation vs. corresponding elastic strain calculations, and **Appendix B.3** for details on the use of completeness thresholding to separate differently deformed regions in the nf-HEDM grain map reconstructions (in this case, we separate twinned from not twinned regions).

## 3. Results
### 3.1 Localized deformation bands

Selected frames from DIC analysis for sample 1 in **Fig. 5** show the 2D surface strains of one of the four faces of the 1×1×8 mm³ gage section using a uniform 10% maximum strain scale bar (**Fig. 5A**) and narrower frame-specific scale bars to contrast local heterogeneities (**Fig. 5B**).

The white boxes in **Fig. 5** outline the portion of the gage section that was illuminated by X-rays during ff-HEDM measurements. The DIC frames corresponding to ff-HEDM measurements are labeled by the load step numbers at the top of each frame, correlating with **Fig. 2A**. The DIC frames show an LDB initiating between 0.5% and 1.4% macroscopic strain (**Fig. 5B**), banding across the gage cross section around 1.4%, propagating through the gage section until the maximum macroscopic strain of 9.7%, and then remaining as sample 1 was unloaded from 9.7% to 9.6%.

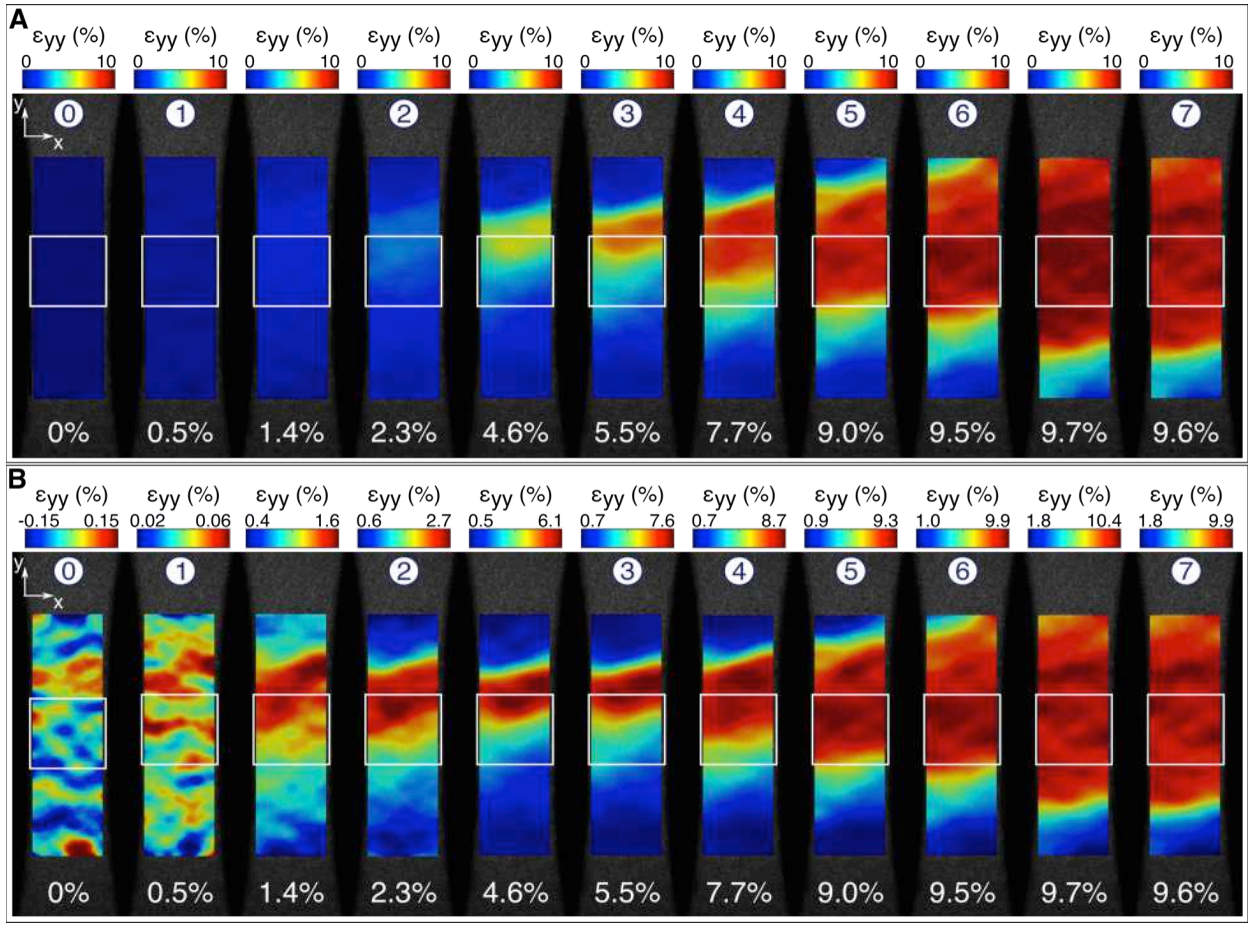

**Fig. 5.** Selected frames from the DIC strain analysis for sample 1. The strain is colored using the scale bars indicated above each frame, with the frames shown in (A) using a uniform scale of 0%–10% strain and the frames in (B) using varying scales to contrast strain variations. The white boxes indicate the region that was illuminated by X-rays during ff-HEDM measurements, and the mean strain in the loading direction in the illuminated region is given at the bottom of each frame. The circled numbers indicate the frames at load steps 0–7.

The DIC frames for samples 2 (**Fig. C.1**) and 3 (**Fig. C.6**) show consistent behavior to sample 1 with an LDB initiating, banding, propagating through the gage section, and then remaining during unloading. Due to the different specimen geometry of sample 3 than that of samples 1 and 2, the LDB remained within the illuminated region of the gage section where the load was concentrated.

**3.2 Variant evolutions with loading**

The stress-strain response of samples 1, 2, and 3 are shown in **Fig. 2**, where the strain reported is the mean surface strain measured from the DIC surface strains corresponding only to the 1×1 mm$^2$ areas of the gage sections that were illuminated by the X-rays (i.e., within the white boxes in **Fig. 5** for sample 1, **Fig. C.1** for sample 2, and **Fig. C.6** for sample 3). The stress-strain curves exhibit three different regions: an apparently elastic initial loading region (blue, marked "I"), a transition to nonlinear behavior followed by a stress drop (red, marked "II"), and a plateau region (green, marked "III").

**Fig. 6A** shows the evolution of the monoclinic texture (within the illuminated volume) of sample 1, where the texture is shown in equal area projection inverse pole figures (IPFs) with respect to the loading axis and the colors correspond to relative volume. The schematic at the top of **Fig. 6A** indicates how the orientations of each of the 12 variants are "clustered" in orientation space (see **Appendix A.3.2**). (Transformation matrices for the 12 variants can be found in, e.g., (Bhattacharya, 2003; Hane and Shield, 1999).) The relative volume fraction of each of the 12 variants were binned at each load step and are shown in the histograms in **Fig. 6B**, where the arrows indicate whether the volume of each variant increased or decreased from the previous load step. At the start of region III (green background), only three variants remain in the microstructure: variants 3, 4, and 9. Following the twin pair definitions in (Bhattacharya, 2003; Hane and Shield, 1999), this combination of variants corresponds to two twin modes: Mode A compound twins consisting of variants 3 and 4, and Mode D Type I/II twins consisting of variants 3 and 9. A closer examination of the reflection symmetry of the variant clusters (see Fig. 4 in (A. Stebner et al., 2013)) showed that these twins are (001) compound twins composed of variants 3 and 4, and ($\bar{1}11$) Type I twins composed of variants 3 and 9. Notice that the relative volume of variant 4 is 0% at load step 2 and > 0% at load step 3. These changes of variants vanishing and then reappearing provide evidence for detwinning as well as twin nucleation, discussed in **Section 4**. At the end of the plateau in region III (load step 6), the microstructure has fully reoriented to the favorable variant, variant 3. Variant 3 is the favorable variant for the grain orientations (in the B2 austenite sense) present in sample 1 because it maximizes the deformation (i.e., work output) in the loading direction (see definitions in **Section 1**).

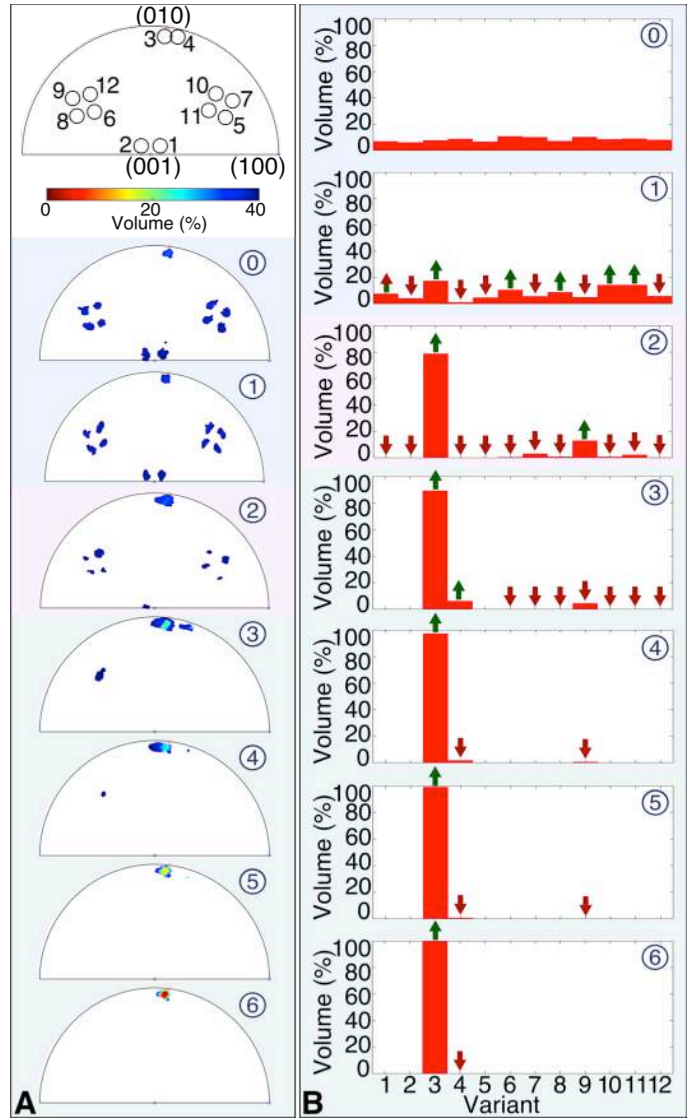

**Fig. 6.** Texture evolution of sample 1. (A) The texture is shown at each load step colored by volume. (C) The volume of each variant at each load step is shown in histograms, with arrows indicating whether the volumes increased or decreased from the previous load step.

The texture evolutions for samples 2 (**Fig. C.2**) and 3 (**Fig. C.7**) exhibit similar behavior to sample 1 with details differing due to different grain (in the B2 austenite sense) orientations and, in the case of sample 3, different sample geometries.

**3.3 Spatially resolved orientation map of the favorable variant at the end of stress plateau**

For sample 1, the nf-HEDM measurement made at load step 6 (see **Table 1**) enabled spatially resolved measurements of the orientations inside the illuminated volume (see **Appendix B**). As seen in **Fig. 6B**, all of the illuminated microstructure had fully reoriented to the favorable variant at load step 6. The reconstruction in **Fig. 7B** and **Movie S1** reveals the internal grain network, where each grain consists only of the fully reoriented favorable variant 3, as well as some intragranular deformation best depicted by the small misorientations within each grain of the 2D misorientation map slices. The colors of the orientations correspond to (*hkl*) as shown in the "Reconstructed Orientations" IPF in **Fig. 7C**. A comparison of the "Orientations Present" (colored by relative volume) and the "Reconstructed Orientations" (colored by (*hkl*)) in **Fig. 7C** show that all of the orientations present in the microstructure have been reconstructed in **Fig. 7B**.

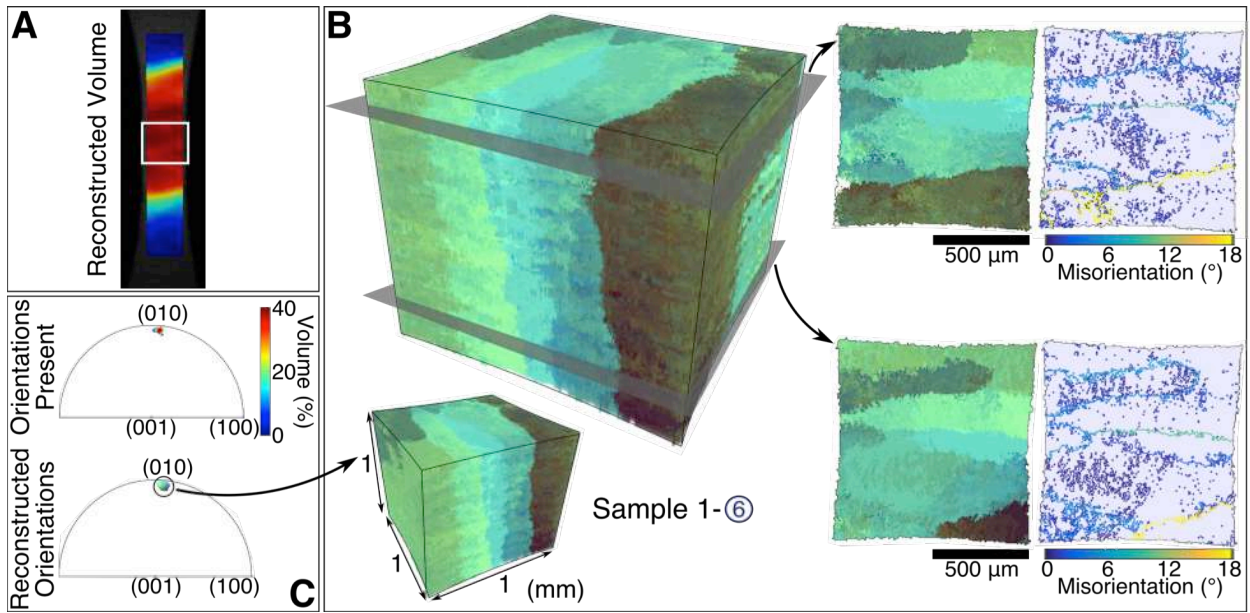

**Fig. 7.** 3D, in situ reconstruction of the detwinned microstructure inside the LDB for sample 1 at load step 6. The 1×1×1 mm$^3$ illuminated volume during the measurement is indicated by the boxed region in (A). The 3D orientation map of the fully reoriented favorable variant is shown in (B), where the colors indicate orientation according to the "Reconstructed Orientations" IPF colormap shown in (C). In (C), the "Orientations Present" IPF shows all of the orientations present in the illuminated volume, colored by relative volume, and the "Reconstructed Orientations" IPF shows all of the orientations reconstructed in (B) colored by (*hkl*).

The nf-HEDM reconstruction for sample 2 (**Fig. C.3**) also shows the full reconstruction of the illuminated volume, where the entire volume consists only of the fully reoriented favorable variant. In contrast, a large volume of the microstructure in sample 3 had fully reoriented to the favorable variant at load step 4, but there are also still twins composed of variants 2, 6, and 10 (see **Fig. C.7**) due to the shorter gage section of sample 3. As discussed in **Appendix B.3**, completeness thresholding of the nf-HEDM data analysis allows separation of twinned regions from not twinned (i.e., fully reoriented) regions. Using this technique, the 3D reconstruction of the illuminated volume in **Fig. 8B** and **Movie S2** is a spatially resolved orientation map of *only* the fully reoriented favorable variant in sample 2. Notice that the reconstructed microstructure crosses through the illuminated volume in the same way that the LDB crosses through the illuminated volume. By reconstructing only the fully reoriented portion of the microstructure, we have shown that the high-strain portion of the LDB consists of the fully reoriented favorable variant and have effectively reconstructed the internal geometry of the LDB itself.

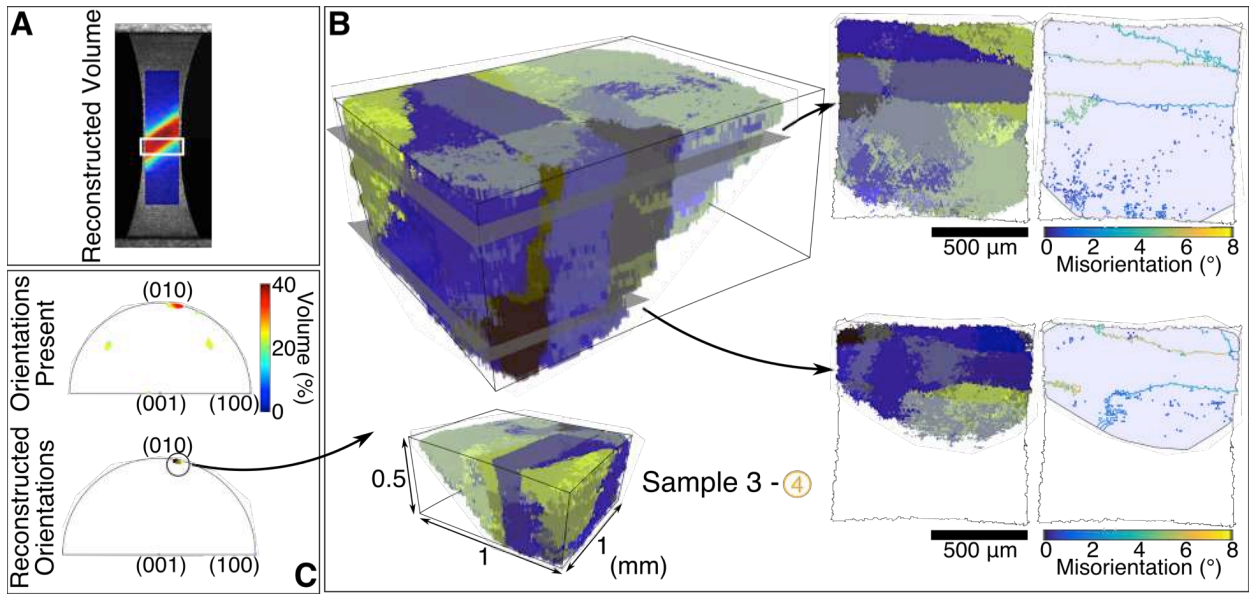

**Fig. 8.** 3D, in situ reconstruction of the detwinned microstructure inside the LDB for sample 3 at load step 4. The 1×1×0.5 mm$^3$ illuminated volume during the measurement is indicated by the boxed region in (A). The 3D orientation map of the fully reoriented favorable variant is shown in (B), where the colors indicate orientation according to the "Reconstructed Orientations" IPF colormap shown in (C). In (C), the "Orientations Present" IPF shows all of the orientations present in the illuminated volume, colored by relative volume, and the "Reconstructed Orientations" IPF shows all of the orientations reconstructed in (B) colored by (*hkl*).

## 3.4 Lattice rotations and elastic strain heterogeneities of the favorable variant inside three different grains

The ff-HEDM data sets were used to make subgrain-resolved measurements of the lattice rotation and corresponding elastic strain experienced by the favorable variant inside three different grains. **Fig. 9** shows the progression of four Bragg reflections, (100), (10$\bar{1}$), ($\bar{1}\bar{2}$1), and ($\bar{1}$21), during loading in sample 1, where the axes $\omega$ and $\eta$ are the diffraction angles defined in **Fig. 4**. Each reflection corresponds to the favorable variant within a specific grain (**Fig. 9B**). Because the grains are slightly misoriented, the reflections are also slightly misoriented (in this case in the $\omega$ direction). At load step 0, the reflections are somewhat "spread," or broad—this is the diffraction signature of small lattice rotations or curvature, likely caused by twin-twin interactions and/or size effects from the undeformed hierarchically twinned microstructure (see **Fig. 3**). By load step 2, the reflections are strongly spread in a particular direction (in this case the $\eta$ direction); this is the diffraction signature of lattice rotations about a specific misorientation axis. This spreading increases in load steps 3 and 4, then suddenly vanishes in load step 5, prior to unloading.

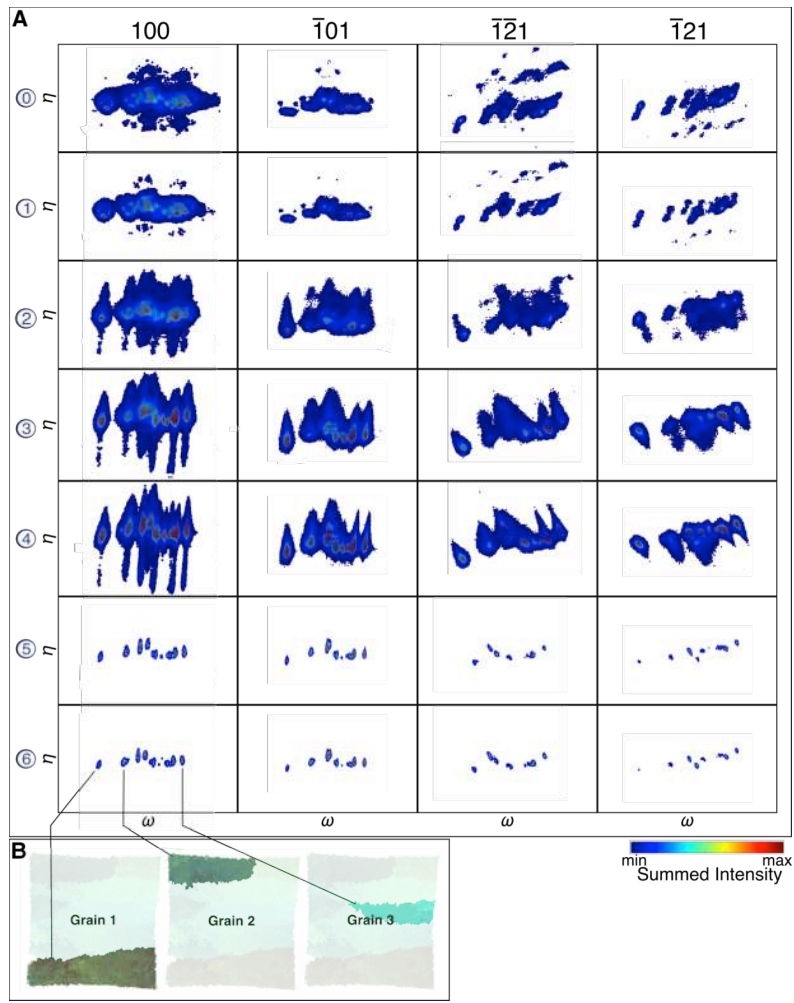

**Fig. 9.** (100), (10$\bar{1}$), ($\bar{1}\bar{2}$1), and ($\bar{1}$21) Bragg reflections from the favorable variant in sample 1 at load steps 0–6 (A), where each reflection corresponds to the favorable variant within a specific grain (B).

The lattice rotation and corresponding elastic strain causing this spreading can be quantified (see **Appendix A.3.3** for details). To explain these measurements, one of the (100) Bragg reflections from **Fig. 9A** is shown again in **Fig. 10A**. Again, this reflection corresponds to the favorable variant 3 in a specific grain and is exhibiting strong spreading in a particular direction. **Fig. 10B** shows the lattice misorientation and corresponding $\varepsilon_{100}$ elastic strain associated with the spreading. The color of each data point indicates the volume of the microstructure undergoing different degrees of misorientation and strain. For example, the left-most data point, which corresponds to the 'top'-most part of the reflection in **Fig. 10A**, indicates that roughly 2% of the lattice is misoriented (relative to the average orientation of the favorable variant in this grain at this load step) by −7°, and this same 2% of the lattice is also elastically strained (relative to the average, unloaded lattice of the favorable variant in this grain) by 0.03%.

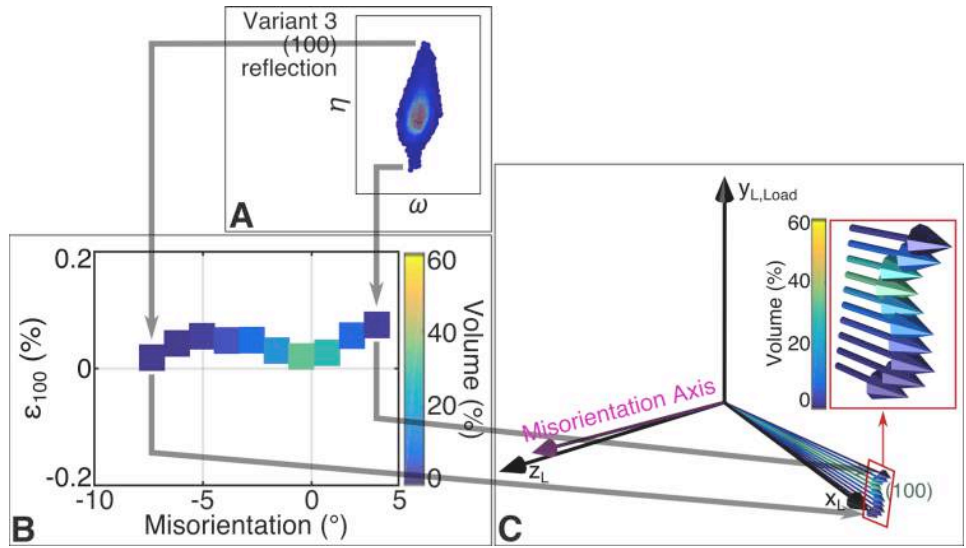

**Fig. 10.** A (100) Bragg reflection from **Fig. 9A** is shown in (A), showing strong spreading in a particular direction. The (100) elastic lattice strain vs. misorientation causing this spreading is shown in (B), where each data point is colored by relative volume. The misorientation axis, nominal (100) normal, and the (100) normals from the most curved portions of the lattice are shown in (C).

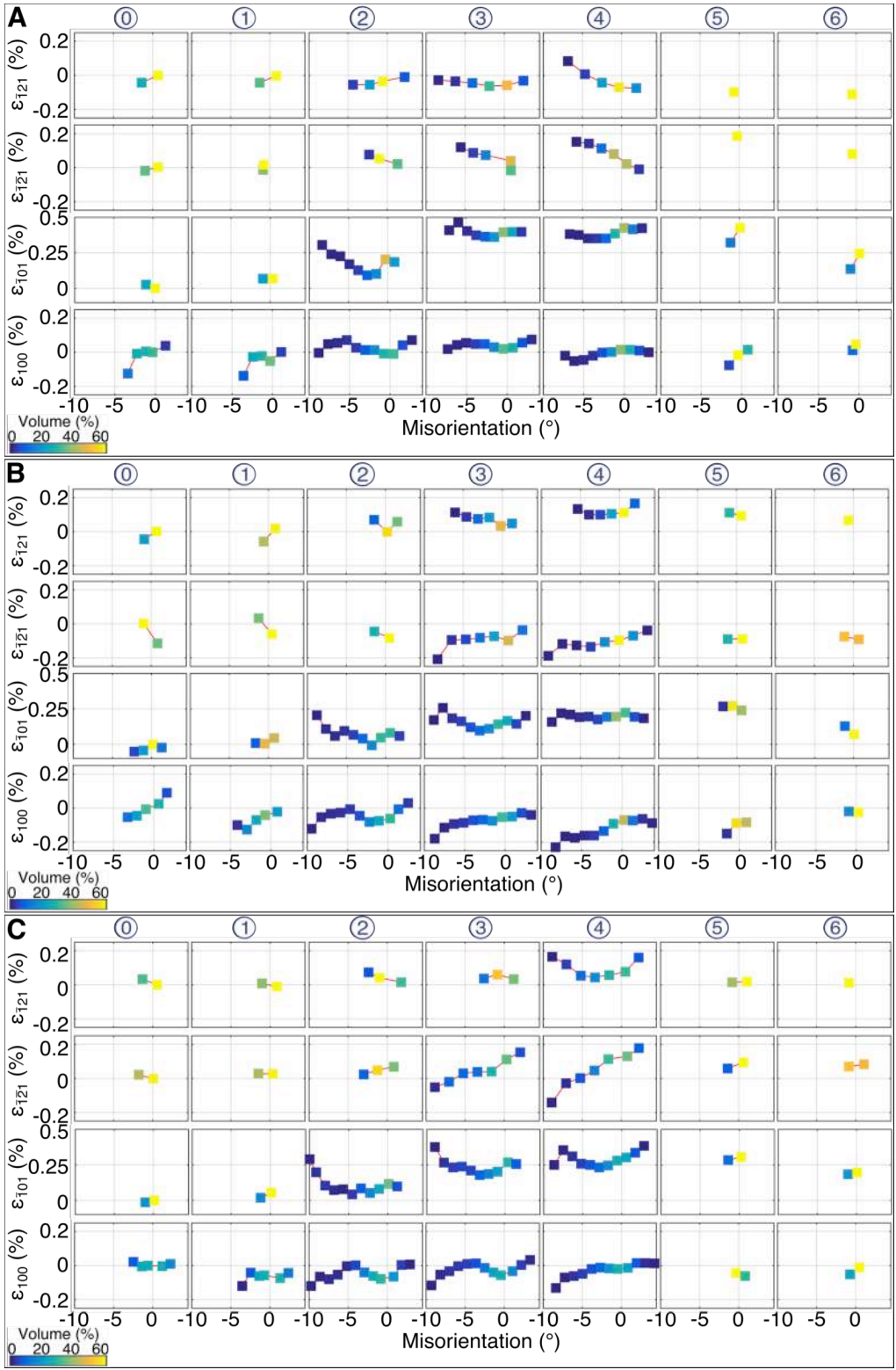

**Fig. 11.** Elastic lattice strain vs. misorientation experienced by the favorable variant within grain 1 (A), grain 2 (B), and grain 3 (C) calculated from the (100), (10$\bar{1}$), ($\bar{1}\bar{2}$1), and ($\bar{1}$21) Bragg reflections.

**Fig. 10C** shows the loading direction, the misorientation axis, and the (100) normals corresponding to the data points in **Fig. 10B**. The nominal (100) normal is considered to be the normal that corresponds to the majority of the lattice, and the normals that deviate from the nominal normal direction correspond to the misoriented lattice. The negative and positive values of misorientation refer to opposite signs of rotation about the misorientation axis, where 'negative' and 'positive' are chosen arbitrarily. The lengths of the normals illustrate the corresponding elastic strains.

The strain-misorientation profiles explained in **Fig. 10** were calculated for all four Bragg reflection types shown in **Fig. 9A** for the three grains shown in **Fig. 9B**. The resulting elastic lattice strain vs. misorientation profiles are shown in **Fig. 11**. The results show that the entire lattice is misoriented by as much as ~12° during deformation, and the lattice rotation occurs about a near-[001] axis (in the laboratory coordinate system) in all three grains (see **Table 2**). For reference, the loading direction, misorientation axes, and nominal plane normals for these four Bragg reflections are shown for the three grains in **Fig. 12**. As expected, the plane normals that are most perpendicular to the misorientation axis (the $(100)$ and $(10\bar{1})$) exhibit the most rotation.

**Table 2.** Misorientation magnitudes and axes for each grain included in the misorientation-lattice strain analysis for sample 1. The axes ($t$) are reported in the laboratory coordinate system shown in **Fig. 4,** as are the ($hkl$).

|  | Misorientation, $\varphi$ (°) | Misorientation Axis, $t$ | Misorientation Axis (hkl) |
|---|---|---|---|
| Grain 1 | 11.59 | [0.18 0.16 0.97] | (0.11 1 $\overline{0.06}$) |
| Grain 2 | 9.98 | [0.06 0.15 0.99] | (0.00 1 $\overline{0.04}$) |
| Grain 3 | 11.64 | [$\overline{0.13}$ 0.11 0.99] | (0.01 1 $\overline{0.06}$) |

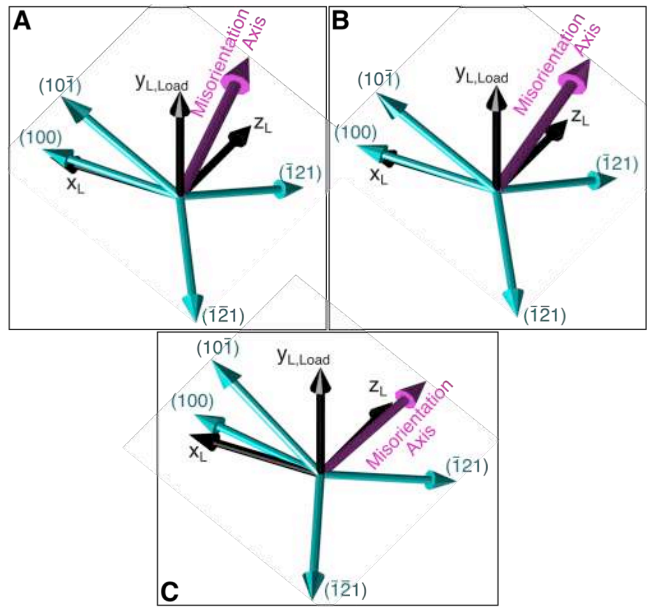

**Fig. 12.** The loading axis, misorientation axis, and nominal (100), (10$\overline{1}$), ($\overline{1}\overline{2}$1), and ($\overline{1}$21) plane normals for the favorable variant in grain 1 (A), grain 2 (B), and grain 3 (C).

The ff-HEDM data sets for sample 2 (**Fig. C.4**) were also used to make subgrain-resolved measurements of lattice rotation vs. corresponding elastic strain experienced by the favorable variant inside three different grains (**Fig. C.5**). The results show that, like sample 1, the entire lattice of the favorable variant in sample 2 is misoriented by as much as ~13°, and the lattice rotation also occurs about a near-[001] axis in all grains (**Table C.1**).

## 4. Discussion
### 4.1 Strain localization initiates via detwinning and twin nucleation, then saturates via detwinning

While this mechanistic insight is not new in and of itself, the data presented in **Section 3** provide detailed, direct observations of the interactions between strain localization and twin nucleation and detwinning. They also provide quantified, variant-by-variant volume fraction evolution measurements through the volume of bulk samples during reorientation processes. Hence, they are valuable for verifying crystal mechanics models of reorientation processes, especially in 3D. Specifically, previous ex situ microstructure studies were able to quantify twin structures with no knowledge of exactly how they formed, while the previous bulk in situ diffraction studies could identify the dominant reorientation mechanisms but lacked knowledge of the twin structures. The ability to perform variant-by-variant analyses with 3DXRD allows for both reorientation mechanisms and twin structures to be simultaneously observed and quantified. Furthermore, the open question of why twin nucleation occurs in some crystals while detwinning dominates others is answered.

#### 4.1.1 Region I: "elasticity + reorganization of existing twins"

In this study, we define region I by two macroscopic behaviors: (1) before the LDB initiates and (2) before the onset of nonlinearity in the stress-strain response. For sample 1, region I includes load steps 0–1 (**Fig. 2A** and **Fig. 5**). Where one might assume that this consists of purely elastic behavior, a comparison of the variant volumes at load steps 0 and 1 (**Fig. 6B**) show that there is also a "shuffling" of the relative volume fractions of the self-accommodation twins. This shuffling, or reorganization, confirms previous observations of reorientation events occurring before the stress plateau (Dilibal, 2013; Liu et al., 2000; A. P. Stebner et al., 2013; Tadayyon et al., 2017). The reorganization seems somewhat random, as the favorable variant 3 is not yet significantly dominating the microstructure, and variants 4 and 9, which make up the dominant twin systems during the plateau, actually decrease in volume between load steps 0 and 1. For this reason, we can infer that this initial reorganization represents reorientation events that have high mobility relative to the events occurring during the stress plateau, similar to highly mobile twin boundary movement observed in Ni-Mn-Ga alloys (Seiner et al., 2014; Straka et al., 2011).

For samples 2 and 3, we did not record any HEDM measurements at load steps in region I, but the results are in agreement with the region I discussion of sample 1. Load step 1 for sample 2 (**Fig. 2B**) is just after region I, and the volumes of the variants are substantially different from the unloaded microstructure (**Fig. C.2B**). Load step 1 for sample 3 (**Fig. 2C**) is also just after region I, and the volumes of each variant has either increased or decreased from that of load step 0 (**Fig. C.7B**).

### 4.1.2 Region II: "detwinning + twin nucleation"

We define region II by two macroscopic behaviors: (1) the initiation of the LDB and (2) the transition to nonlinear behavior and stress drop in the stress-strain response. For sample 1 (**Fig. 2A**), region II starts just after load step 1 and ends just before load step 3. **Fig. 5** shows the LDB initiating and banding between load steps 1 and 3, but the strain inside the LDB has not yet saturated to its peak value. **Fig. 6B** shows that the volume of the favorable variant 3 increases from less than 20% to nearly 75%, much of which likely formed via detwinning as has been previously observed in region II (Tadayyon et al., 2017). Twin nucleation is also occurring during region II, as revealed by the volume of variant 4 that disappears by step 2 and then reappears by step 3. These nucleated twins are (001) compound twins consisting of variants 3 and 4, the same twin system frequently observed to form from [011] Type II twins early in stress-induced reorientation of monoclinic NiTi (Liu et al., 2000). After load step 2, there is no further evidence of twin nucleation (**Fig. 2C**) corresponding to when the (light blue, ~4% in **Fig. 5A**) low-strain gradient region of the LDB is no longer within the illuminated volume. For this reason, we can infer that the low-strain gradient portion of the LDB contains detwinning plus twin nucleation.

While the reason for twin nucleation in reorientation processes has yet to be fully understood, these new data suggest that twin nucleation occurs because it is physically necessary to fully reorient to the favorable variant. **Fig. 13** depicts all of the possible pathways from variants 1–2, 4–12 to the favorable variant 3 for sample 1 using only compatibility-maintaining twinning operations (Bhattacharya, 2003; Hane and Shield, 1999). Variants 5, 8, 10, and 12 cannot form twins with the favorable variant 3 and would theoretically require an intermediate twin system to reorient to variant 3. For example, variants 5, 8, 10, and 12 could reorient to variant 4 and then to the favorable variant 3. The necessity of these intermediate twin systems explains the tendency of new nucleate twins early in the loading process. For example, the

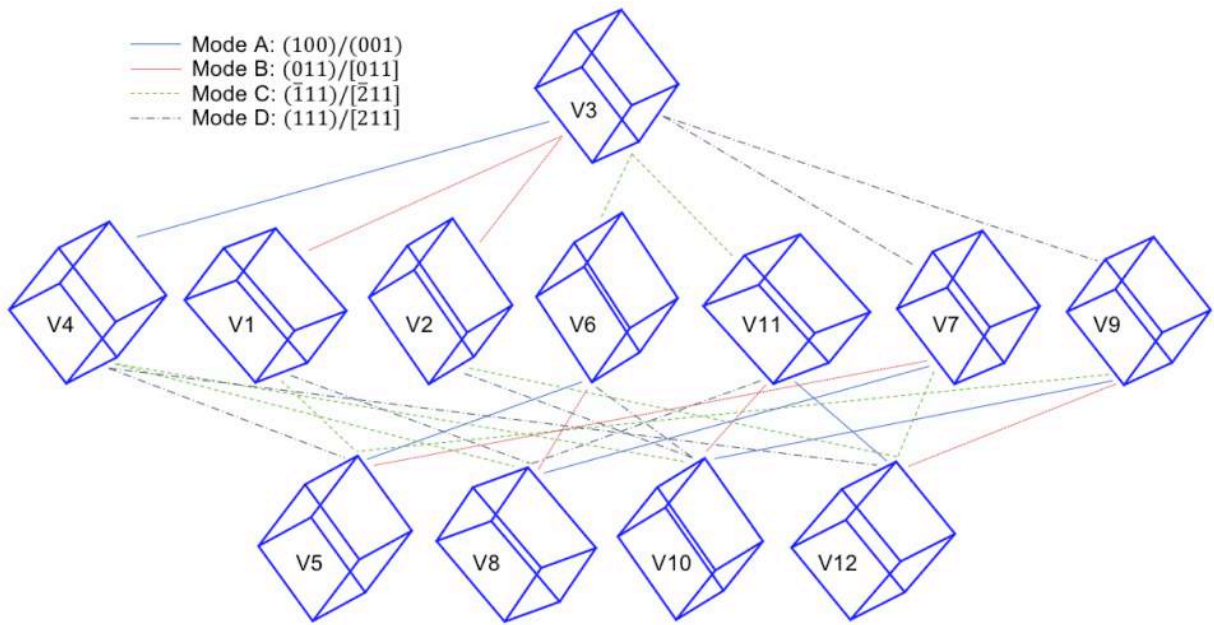

**Fig. 13.** Ferroelastic twin relationships between the preferred variant 3 (V3) for sample 1 and the other 11 variants (V1, V2, V4–V12). The twin mode designations follow (Hane and Shield, 1999), where the (*hkl*)/[*uvw*] of the twin-plane-normal/twin-shear-direction are given with respect to monoclinic crystallography for the Type-I/ II twins, respectively, except for the compound twins of Mode A where the twin planes are rational for both twin types. Only the first and second order connections to variant 3 are shown. Other twin pairs exist across the rows of this figure but are omitted for legibility.

necessity of the pathway from variants 5, 8, 10, and 12 to the favorable variant 3 via variant 4 may be the reason that the (001) compound twins consisting of variants 3 and 4 appear to nucleate in load step 3 in **Fig. 6**

Even with these X-ray techniques, twin nucleation is challenging to identify. It is somewhat fortunate that the volume of variant 4 decreased so much before increasing substantially so that we could definitively identify it as forming via twin nucleation. Another probable observation of twin nucleation is for variant 9 in sample 1 between load steps 1 and 2 (**Fig. 6B**). Variant 9 can be seen decreasing to a very low volume at load step 1 before suddenly increasing substantially to the second most dominant variant in load step 2. In sample 2, variants 5 and 12 also increased in volume substantially between load steps 0 and 1 (**Fig. C.2B**). In these latter examples, however, it is not possible to say whether such events occurred via detwinning only or detwinning plus twin nucleation.

**4.1.3: Region III: "detwinning"**

We define region III by two macroscopic behaviors: (1) the propagation of the strain-saturated LDB and (2) the stress plateau. By region III for all three samples, there are three variants remaining: the favorable variant dominating the microstructure plus two additional variants in lower volume fractions. For sample 1 at load step 3 (**Fig. 2A**), there remains the favorable variant 3 plus variants 4 and 9 (**Fig. 6B**). These variants correspond to (001) compound twins consisting of variants 3 and 4 and ($\bar{1}11$) Type I twins consisting of variants 3 and 9. Region III deformation is driven by these remaining twin systems detwinning to the favorable variant until the microstructure has fully reoriented to the favorable variant. These detwinning processes occur during load steps 3–6 (**Fig. 6B**) and are nearly complete by load step 5. Correlating the variant volume fractions in **Fig. 6B** to the DIC analysis in **Fig. 5**, the medium-strain (yellow, ~6% in **Fig. 5A**) portion of the LDB gradient moves through the illuminated volume during load steps 3–6, and is almost completely through the illuminated volume by load step 5. Thus, we can infer that the detwinning occurs in the medium-strain gradient portion of the LDB, and the strain-saturated (red, ~9% in **Fig. 5A**) region is the fully reoriented favorable variant.

These findings are consistent for samples 2 and 3. For sample 2 at load step 2 (**Fig. 2B**), there remains the favorable variant 4 plus variants 5 and 12 (**Fig. C.2B**), corresponding to ($\bar{1}11$)

Type I twins composed of variants 4 and 12, and [$\bar{2}$11] Type II twins composed of variants 4 and 5. These two twin systems detwin during load steps 2–4 (**Fig. C.2B**) as the medium-strain (yellow, ~6% in **Fig. C.1A**) gradient portion of the LDB moves through the illuminated volume. For sample 3 at load step 2 (**Fig. C.7B**), there remains the favorable variant 2 plus variants 6 and 10, corresponding to ($\bar{1}$11) Type I twins composed of variants 2 and 6, and [$\bar{2}$11] Type II twins composed of variants 2 and 10. Because the specimen geometry was different for sample 3 (**Fig. C.6**), the strain-saturated portion of the LDB never completely filled the illuminated volume and twins remained in the gradient portions of the LDB (**Fig. C.7B**). As a result, there is a very small volume of variant 11 detected in the regions outside of the LDB in the illuminated volume at load step 2.

The spatially resolved orientation maps confirm the conclusion that the strain-saturated region of the LDB contains the fully reoriented favorable variant, and the gradient region of the LDB contains twins. For sample 1 (**Fig. 7**), the strain-saturated portion of the LDB had fully passed through the illuminated volumes at the time of the nf-HEDM collection, so the reconstruction revealed the full grain network where each grain consists only of the favorable variant (i.e., no twins). For sample 3 (**Fig. 8**), the strain-saturated portion of the LDB never completely filled the illuminated volume, so the reconstruction revealed the grain network only inside the fully reoriented favorable variant inside the strain-saturated portion of the LDB. Outside of the strain-saturated portion of the LDB (i.e., in the gradient region), the material still contained twins.

**4.2 Strain localization leads to geometrically necessary elastic lattice rotations and strains**

**Fig. 11** shows that the favorable variant domains in sample 1 are elastically misoriented or curved by as much as 12° during load steps 2–4, that an S-shaped relationship exists between lattice misorientation and elastic strain, and that this misorientation almost completely disappears by load step 5, prior to unloading. Examining the DIC frames in **Fig. 5**, load steps 2–4 are when the gradient regions of the LDB are passing through the illuminated volume. When **Fig. 5** shows that the gradient region of the LDB has almost completely passed through the illuminated volume at load step 5, **Fig. 11** shows the lattice misorientation has disappeared (i.e., the lattice has relaxed). The results for sample 2 also show that the lattice of the favorable variant is misoriented by ~13°, that there is an S-shaped relationship between lattice rotation and elastic strain (**Table C.1**), and that this misorientation relaxes after the gradient region of the LDB has passed through the illuminated volume (**Fig. C.1**, **Fig. C.5**). Notice that this mechanism occurs in both samples 1 and 2, even though the two samples consisted of different crystallographic orientations, favorable variants, and twin systems.

This misorientation most likely occurs across the gradient region of the LDB due to geometric constraints dictated by necessary macroscopic geometric continuity. In other words, inside the strain-saturated region of the LDB, the sample has undergone large macroscopic deformation (elongation in the loading direction and cross-section reduction) via twin reorientation, while outside the this region, the sample is not as macroscopically deformed. Because a cleavage plane did not develop, the lattice must then curve, or rotate, across the gradient region of the LDB to transition from the highly deformed to less deformed gage geometry. This curvature is the reason for the S-shaped lattice rotation and elastic strains (discussed more in **Section 4.3**). Because the misorientation only exists across the gradient regions of the propagating LDBs, the relative volume of the amount of material that is misoriented does not change appreciably while the misorientation is present (load steps 2–4 in **Fig. 11** for sample 1). Furthermore, because the angle of the LDBs does not change with loading, the misorientation axis is consistent through all load steps within regions II and III, even when the texture changes between load steps.

For both sample 1 (**Table 2**) and sample 2 (**Table C.1**), the axis about which the lattice is rotated is near-[001] in terms of the laboratory coordinate system and near-(010) in terms of (*hkl*) of the favorable variants. The (010) axis is not a twin plane normal. It happens to lie in the plane of the compound twins in sample 1 but does not lie within the twin planes of the twins present in sample 2 for load steps 2–4. This evidence indicates that the misorientation is not primarily due to elastic interaction between the twins themselves or their interfaces. **Fig. 14** shows that the misorientation axes align with the traces of the LDB edges in the DIC frames for both sample 1 (**Fig. 14A**,**B**) and sample 2 (**Fig. 14C**,**D**). We hypothesize that the misorientation occurs about an axis that exists at the intersection of the LDB plane and the surface normal, that this is the reason that both axes are near-[001] in terms of the laboratory coordinate system, and that this is why the projection of the misorientation axis aligns with the LDB edges in the DIC frames. The fact that both misorientation axes are near-(010) in terms of (*hkl*) of the favorable variants may just be a coincidence.

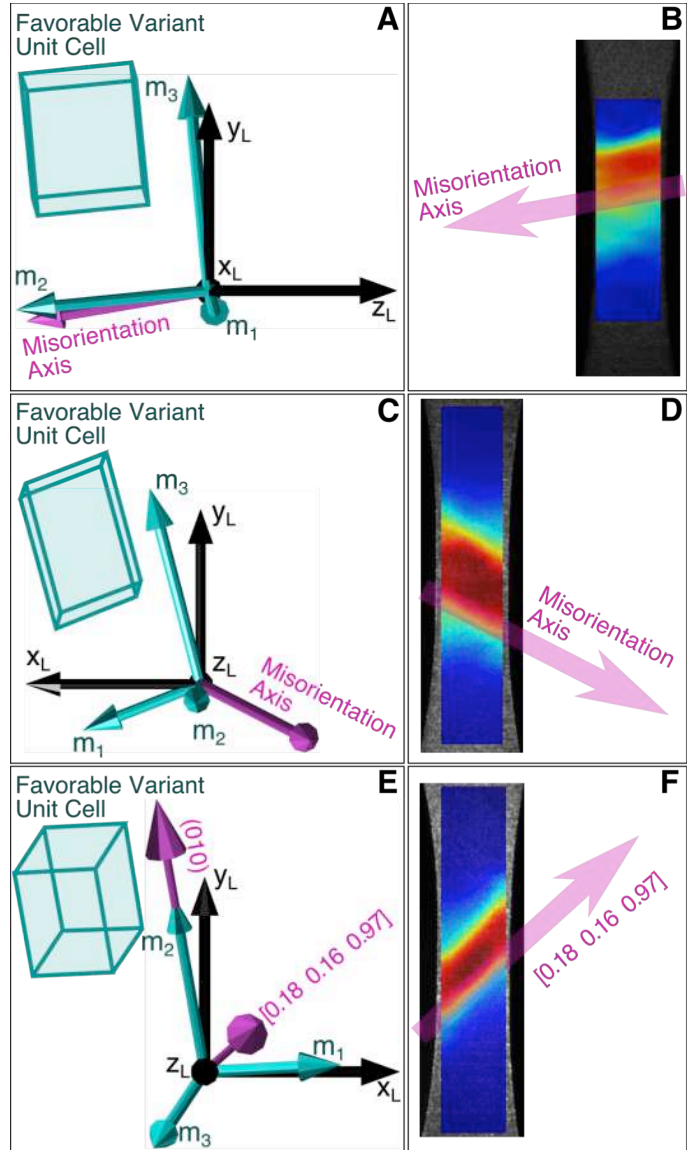

**Fig. 14.** The misorientation axes are shown from the viewing angles of the DIC camera relative to the lattice vectors of the favorable variants and the loading direction $y_L$ for samples 1 (A) and 2 (C). The misorientation axes are shown again on top of DIC frames for samples 1 (B) and 2 (D). The (010) normal and a [0.18 0.16 0.97] vector (in terms of the laboratory coordinate system) is shown for sample 3 (F) and is shown on top of one of a DIC frame (F).

Sample 3 provides an opportunity to test this hypothesis. While the data were not sufficient for the same detailed misorientation analyses, the orientation of the favorable variant for sample 3 is known, as is the relative orientation of the LDB in the DIC data. **Fig. 14E,F** shows the visualization of these orientation relationships. Unlike samples 1 and 2, the (010) normal is significantly misaligned from the LDB; in fact, it is nearly aligned with the loading axis, such that misorientations about (010) normal would not accommodate strain localization.

If we include one of the near-[001] vectors in terms of the laboratory coordinate system ([0.18 0.16 0.97] from sample 1, **Table 2**), then this global vector again aligns with the LDB edge in the DIC frame. One discrepancy is why the misorientation vector is perpendicular to two of the sample faces (the $\pm z_L$ faces) but not the other two faces (the $\pm x_L$ faces), but it is important to note that these are only the *dominant* modes of misorientation and that other, lesser modes were not measured. Note: this discussion does not attempt to explain the different LDB orientations in the three samples.

**4.3 There are multiple similarities between twin rearrangement deformation bands and plastic slip shear bands**

The elastic lattice rotations and strains observed during ferroelastic twin reorientation are analogous to the lattice rotations across the edges of shear bands and necking regions observed during plastic deformation of ductile single crystals (Elam, 1927; Fleischer and Chalmers, 1957; Peirce et al., 1983, 1982). In fact, there are several similarities between reorientation LDBs and shear bands, including the degree of the lattice rotation (5–30° is reported in (Elam, 1927; Fleischer and Chalmers, 1957; Peirce et al., 1982)), the S-shaped misorientation-strain profiles across the band edge (Peirce et al., 1982), and the frequent existence of two or more slip systems (or in this case twin systems) during banding (Peirce et al., 1983, 1982). Multiple slip systems are often necessary for geometric stability during shear band formation under fixed grip conditions (Fleischer and Chalmers, 1957), often occurring as two primary-conjugate slip systems where one is considered to be a minor slip system. This may be the reason that we consistently see two or three twin systems during the detwinning process within the LDBs (**Fig. 6B**, **Fig. C.2B**, and **Fig. C.7B**) and even the existence of two near-perpendicular LDBs in some DIC frames for sample 2 (**Fig. C.1** from 2.9–4.3%).

Another important similarity between lattice rotation in reorientation LDBs and shear bands is a geometric softening effect (Chang and Asaro, 1981; Peirce et al., 1983, 1982). In shear bands, geometric softening refers to the increase in RSS on a slip system due to lattice rotation, making it "easier" for that slip system to activate. **Fig. 15** shows the RSS on the (100) and (001) compound twin systems consisting of variants 3 and 4 (the dominant variants present in sample 1 during region III deformation) as a function of the lattice misorientation (**Table 2**). The RSS on the other six systems are shown assuming no lattice misorientation for comparison— most of these twin systems have already vanished by region III deformation of sample 1 and are no longer present during the observed lattice misorientation.

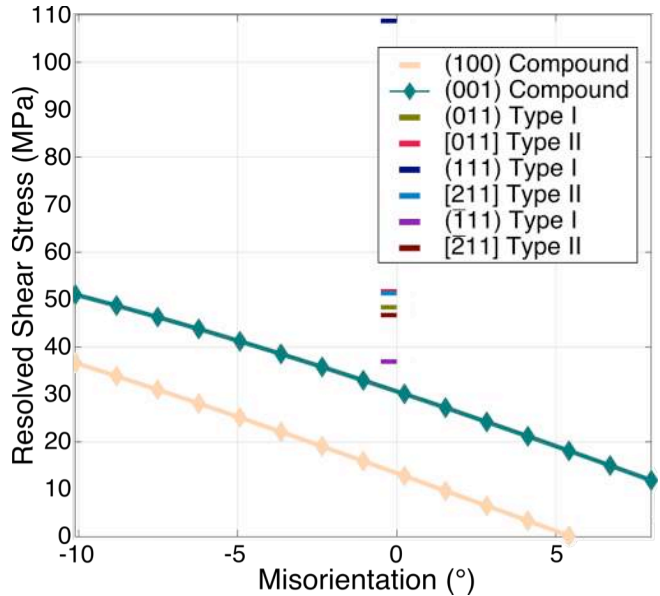

**Fig. 15.** The effect of misorientation on the resolved shear stresses on the compound twin systems are shown relative to the six other twin types for sample 1.

The six reference twin systems have relatively high RSS values of 40–110 MPa. The compound twin systems are oriented such that the twin plane or shear directions are nearly perpendicular to the loading direction, resulting in relatively low RSS values of 15–30 MPa at 0° misorientation. When we include the observed lattice misorientation, we see an increase in RSS on the compound twin systems to RSS values comparable to those experienced by the other six, already detwinned systems. This observation has several implications. First, the geometric softening effect due to lattice rotations may in some cases be necessary for full reorientation to the favorable variant. Second, there is a critical RSS for detwinning to occur for this material system. Third, if detwinning is easiest inside the gradient regions of the LDB due to this geometric softening effect, then this localized detwinning will continually shift the edge of the LDB, thus promoting the propagation of the LDB through the gage section.

## 5. Conclusion

The insights presented in this paper document important micromechanics of ferroelastic twin reorientation enabled by using a suite of modern 3D X-ray diffraction capabilities. In the case of studying load-induced twin reorientation in SMAs, researchers were previously forced to either stitch together 2D snapshots of microscale measurements or infer micromechanics from averaged macroscopic measurements. Here, we have used 3D in situ diffraction techniques to develop a complete picture of the relationships among the twinning mechanisms, the evolution of the twins in the microstructure, and the macroscopic response, all visually summarized in **Fig. 16**.

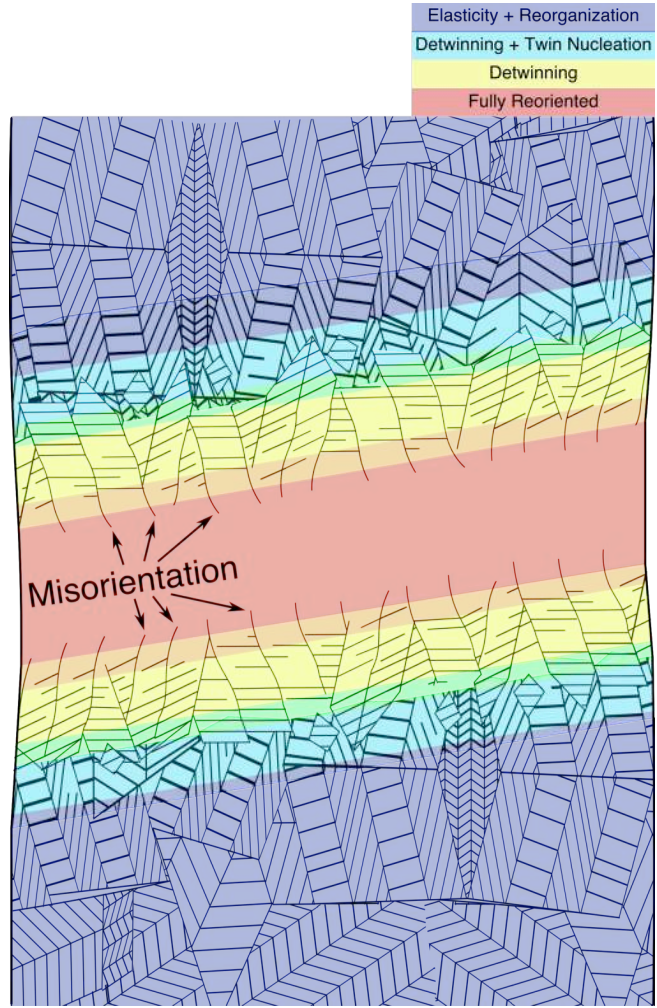

**Fig. 16.** A schematic summarizing the microstructure-based mechanisms underlying load-induced ferroelastic twin reorientation in bulk materials.

- In the early, approximately linear portion of the stress-strain response, prior to the formation of an LDB, the material primarily deforms via elastic deformation with minor initial reorganization of the existing twins (dark blue in **Fig. 16**).

- As the stress-strain behavior becomes nonlinear including a stress drop and the initiation and banding of an LDB, the microstructure is reorienting via detwinning plus twin nucleation. This reorientation produces strains that are a fraction of the eventual saturated values, taking place in the relatively low-strain gradient region of the LDB (light blue in **Fig. 16**).

- As the macroscopic behavior enters the stress plateau, higher strain values, and the strain-saturated region of the LDB propagates through the gage section, the microstructure fully reorients via detwinning only (yellow in **Fig. 16**). The strain saturates as the microstructure fully reorients to the favorable variant (red in **Fig. 16**).

- As the lattice crosses the gradient region of the LDB, the lattice is forced to elastically rotate and strain due to geometric constraints dictated by the macroscopic geometric continuity between the relatively low-strain (dark blue in **Fig. 16**) and high-strain regions (red in **Fig. 16**).

- Through geometric softening, this lattice rotation will affect the ability of the microstructure to fully reorient to the favorable variant and may even be necessary to both full reorientation as well as the propagation of the LDB through the gage section.

These findings confirm insights made by previous researchers as well as introduce newly discovered mechanics of load-induced ferroelastic twin reorientation. By tracking the volume of all variants at different stages in loading, the information presented in this study also provides valuable statistical and crystal-resolved data for informing and verifying future modeling efforts. These discussions will help guide future researchers using ferroelastic twin rearrangement to control macroscopic mechanical, electrical, and magnetic properties in novel technologies using bulk multiferroics.

      3DXRD techniques such as nf- and ff-HEDM have mostly been applied to studying nonferroic behaviors of cubic and hexagonal crystal structures. This work demonstrates the ability for similar 3D, in situ, multiscale studies to elucidate the complex behaviors of low-symmetry, ferroelastic, and/or multiferroic materials that are capable of numerous types of deformation. By measuring across several orders of magnitude in length scales, these kinds of experiments can improve our understanding of complicated material behaviors and provide opportunities to accelerate our abilities to model them. For example, the presence of twin nucleation to maximize deformation under the constraints of kinematic compatibility reveals a

strong interaction between microstructure-scale crystal mechanics and macroscale continuum deformation. In other words, materials with multiple possible modes of deformation will "find a way" to maximize deformation. The geometric softening enabled by lattice curvature across the localized deformation bands is another example of a somewhat surprising, opportunistic path to energy-minimizing responses, and it is also another example of the interplay between micro- and macroscale mechanics. As discussed in **Section 4.3**, these types of observations have strong analogies to other types of material responses, and these types of experiments can inform (and be informed by) a diverse variety of fields. The ability to measure material responses in situ, in 3D, and across length scales over statistically significant volumes is critical to developing material models that capture the operative micromechanics that govern macroscale response.

**Acknowledgements**

ANB acknowledges the support provided by the National Science Foundation Graduate Research Fellowship Program under award DGE-1057607. ANB and APS acknowledge support provided by the National Science Foundation under award CMMI-1454668, Mechanics of Materials and Structures program. LC and MJM acknowledge support provided by the U.S. Department of Energy Office of Science under award DE-SC0001258. This work is based upon research conducted at the Cornell High Energy Synchrotron Source (CHESS), which is supported by the National Science Foundation under awards DMR-1332208 and DMR-0936384. ANB and APS acknowledge XSEDE resources under awards TG-MSS160032 and TG-MSS170002. ANB also acknowledges Jette Oddershede and Margaret Koker for helpful discussions regarding nf-HEDM reconstructions.

## Appendix A. Far-field high-energy diffraction microscopy analyses and visualizations

### A.1 Data collection

Monochromatic X-ray energies were set to the middle of the $K_\alpha$ absorption edges of holmium (55.618 keV) for samples 1 and 2 and ytterbium (61.332 keV) for sample 3 were used. The load steps for ff-HEDM data collection are indicated in **Table 1**. The X-ray beam was vertically centered on the specimen gage section and masked to 1 mm tall × 2 mm wide using slits. A GE41RT amorphous silicon area detector with 2048 × 2048 pixels of 200 × 200 μm² size was used to record diffraction patterns. Exposures were taken at each 0.1° $\omega$ increment of sample rotation (defined in **Fig. 4**) over a full 360° rotation, resulting in 3,600 detector images recorded per ff-HEDM data point. A dark image of the detector (i.e., an exposure without the X-ray beam on) was also taken before and after each sample rotation to characterize the background noise of the detector at the time each measurement was made. The detector distance from the sample, the center position of the beam on the detector, the detector tilts, and detector distortions were characterized by performing least squares refinements of detector calibration parameters within the HEXRD software package (Bernier, 2017) using diffraction measurements of a $CeO_2$ standard powder (NIST RSM 674b) at $\omega = 0°$ and $\omega = 180°$. The ff-HEDM detector calibration parameters are given in **Table A.1**. The calibration parameter names, variables, and application to the data can be found in (Bernier et al., 2011).

**Table A.1.** ff-HEDM detector calibration parameter values.

| Detector calibration parameter | Parameter values for the 55.618 keV experiments | Parameter values for the 61.332 keV experiment |
|---|---|---|
| $x$ position of beam center | −1.20 mm from detector center | −1.60 mm from detector center |
| $y$ position of beam center | 1.60 mm from detector center | 0.40 mm from detector center |
| sample-to-detector distance | 1012.4 mm | 904.6 mm |
| tilt of detector about $x$-axis | 0.064 radian | −0.004 radian |
| tilt of detector about $y$-axis | 0.052 radian | −0.000 radian |

**A.2 Data analysis**

The ff-HEDM data sets were analyzed using the HEDM software suite HEXRD (Bernier, 2017). The martensite phase poses several challenges to conventional ff-HEDM data analysis methods such as those used within HEXRD (Bucsek et al., 2018). These challenges arise from the low (monoclinic) symmetry of the martensite phase, which results in greater numbers of Debye–Scherrer rings within a given $2\theta$ range ($2\theta$ defined in **Fig. 4**) and large structure factor disparities amongst those rings. For example, the (*hkl*) family, *d*-spacing, $2\theta$, and normalized structure factors for monoclinic NiTi shown in **Table A.2** demonstrate these challenges for the first 18 Debye–Scherrer rings recorded on the ff-HEDM detector for the 55.618 keV experiments. For the 55.618 keV experiments with a sample-to-detector distance of 1.2 m, a maximum $2\theta$ of 9.7° can be measured, resulting in a resolution of 0.01° per pixel.

**Table A.2**. (*hkl*), *d*-spacing, 2θ, and normalized structure factors for the first 18 Debye–Scherrer rings present on the ff-HEDM detector during measurements of samples 1 and 2, sorted by increasing 2θ.

| h | k | l | d-spacing (Å) | 2θ (°) | Normalized structure factor |
|---|---|---|---|---|---|
| 0 | 0 | 1 | 4.626 | 2.762 | 14 |
| 0 | 1 | 1 | 3.077 | 4.151 | 1 |
| 1 | 0 | 0 | 2.869 | 4.453 | 3 |
| −1 | 0 | 1 | 2.591 | 4.930 | 5 |
| 1 | 1 | 0 | 2.355 | 5.426 | 20 |
| 0 | 0 | 2 | 2.313 | 5.525 | 53 |
| 1 | 0 | 1 | 2.309 | 5.533 | 0 |
| −1 | 1 | 1 | 2.194 | 5.825 | 92 |
| 0 | 2 | 0 | 2.061 | 6.201 | 50 |
| 0 | 1 | 2 | 2.017 | 6.336 | 100 |
| 1 | 1 | 1 | 2.015 | 6.343 | 0 |
| −1 | 0 | 2 | 1.925 | 6.638 | 3 |
| 0 | 2 | 1 | 1.883 | 6.789 | 2 |
| −1 | 1 | 2 | 1.744 | 7.328 | 15 |
| 1 | 0 | 2 | 1.698 | 7.529 | 1 |
| 1 | 2 | 0 | 1.674 | 7.636 | 1 |
| −1 | 2 | 1 | 1.613 | 7.925 | 2 |
| 1 | 1 | 2 | 1.570 | 8.144 | 0 |

However, given that each martensite reflection can occupy 30 or more pixels across the 2θ direction due to grain size, elastic strains, plasticity, size effects, etc., a distinct (*hkl*) ring should be separated by at least 0.30° (and preferably more) from other (*hkl*) rings on either side.

    HEXRD models the theoretical Bragg reflections that should be observed for a given crystal orientation and then assigns a completeness to each orientation. The *completeness* is the percentage of Bragg reflections identified by HEXRD for each crystal versus the total theoretical number of reflections that could be generated by that crystal. When two or more Debye–Scherrer rings are too close to each other in 2θ, one ring's reflection may mistakenly be identified as the reflection of another ring. Furthermore, the dynamic range of the GE41RT detector is 14-bit, and 2-bits of background noise are typical, reducing the effective dynamic range to 12 bits. This bit limitation makes it challenging to measure usable signal from (*hkl*)s that have large structure factor disparities. If attenuation is set to not saturate the signal from high structure factor (*hkl*)s, the lower structure factor (*hkl*)s become indistinguishable from the background noise; yet when the attenuation is adjusted to distinguish low structure factor (*hkl*)s, the high structure factor (*hkl*) signals saturate, and information about the statistical distribution of diffraction events (such as peak centers) cannot be analyzed. For this reason, only the first four Debye–Scherrer rings were

used for the ff-HEDM analysis using HEXRD, because they are the most well separated in $2\theta$ and also of similar structure factor.

In this work, crystals with a completeness of 55% or more were accepted. This threshold was chosen because when the completeness values were lowered to below 55%, "noise" orientations that were not crystallographically possible variants of martensite were falsely identified. The output files of HEXRD provide the crystal positions, elastic strain tensors, and 3D orientations as well as the reflection locations and intensities observed on the detector that were used to identify each crystal. However, in this work, because of the disparity in both structure factors as well as the sizes of the crystals during most loading steps, we only report crystal orientations and volumes. The volumes were calculated according to the procedure described in **Section A.3.1**.

### A.2.1 Resolved shear stress calculations of twin systems

The resolved shear stresses (RSS) on the twin systems reported in **Fig. 14** were calculated much like the RSS on a slip system would be calculated. Instead of the slip plane, the twin plane is used, and instead of the slip direction, the shear direction is used. This procedure is described in (Zhang et al., 2000). The specific equation for calculating the RSS $\tau$ on a twin system composed of a twin plane normal $\hat{\boldsymbol{n}}$ and a shear direction $\boldsymbol{a}$ from a uniaxial load of a magnitude $\sigma$ in a direction $\hat{\boldsymbol{e}}$ is

$$\tau = \pm\sigma(\hat{\boldsymbol{n}}\cdot\hat{\boldsymbol{e}})(\hat{\boldsymbol{e}}\cdot\boldsymbol{a})/\|\boldsymbol{a}\|\ . \tag{A.1}$$

The definitions of twin plane and shear direction for twin systems can be found in (Hane and Shield, 1999; Bhattacharya, 2003; Bucsek et al., 2016).

### A.3 Data visualization
### A.3.1 Monoclinic inverse pole figures (IPFs)

As mentioned in **Section A.2**, HEXRD outputs the reflection locations and intensities that were used to identify each crystal. As commonly performed in diffraction analyses (e.g., (Sharma et al., 2016)), we can use the relative intensities of the reflections to calculate the relative volume associated with each orientation. We calculated the relative volume $V_i$ of each B19′ monoclinic (martensite) crystal *i* using the intensities *I* that HEXRD reported for the first three B19′ (*hkl*) rings ({100}, {011}, and {001}) using **Equation A.2**:

$$V_i = \frac{I_i^{\{100\}} + I_i^{\{011\}} + I_i^{\{001\}}}{\sum_{i=1}^{n} I_i^{\{100\}} + I_i^{\{011\}} + I_i^{\{001\}}}. \tag{A.2}$$

In the event individual crystals were not observable, **Equation A.2** can instead be used to calculate the percentage of the diffracted volume that consists of an orientation *i* (which may consist of many crystals). The monoclinic inverse pole figures (IPFs) therefore show the orientations present at each load step colored by the relative volume of each orientation. The IPFs are equal area projections of (*hkl*) orientation distributions plotted relative to the loading axis $y_L$ as defined in **Fig. 4**.

### A.3.2 Variant histograms

The variant histograms in **Figs. 6B, C.2B,** and **C.7B** were calculated using the B19′ monoclinic martensite orientations reported by HEXRD. The variant naming convention is the same as in (Bhattacharya, 2003; Hane and Shield, 1999). As discussed in the main body of the paper, the samples were all highly textured and consisted of 5–15 grains, where grains are defined here in the B2 cubic austenite phase. These austenite grains led to martensite variant orientations that were grouped into distinct "clusters" (see **Fig. A.1**). For example, the variant 1 orientations were closely oriented, or "clustered," in orientation space, regardless of which grain the variant resided in. Additionally, the martensite variants were well separated from each other in orientation space. For example, variant 1 orientations were well separated from all variant 2 orientations and so on, regardless of which grain the variants resided in. This variant orientation clustering is illustrated in **Fig. A.1**.

To assign each martensite orientation to a variant, the average B2 cubic austenite orientation needed to be calculated. To do this, ten million B2 orientations were randomly generated within the fundamental B2 (simple cubic) orientation space. For each B2 orientation, we calculated the orientations of all possible B19′ monoclinic martensite twins using the crystallographic theory of martensite (CTM) (Ball and James, 1989; Bhattacharya, 2003; Bilby and Crocker, 1965; Wechsler et al., 1953) and the procedure outlined in (Bucsek et al., 2018).

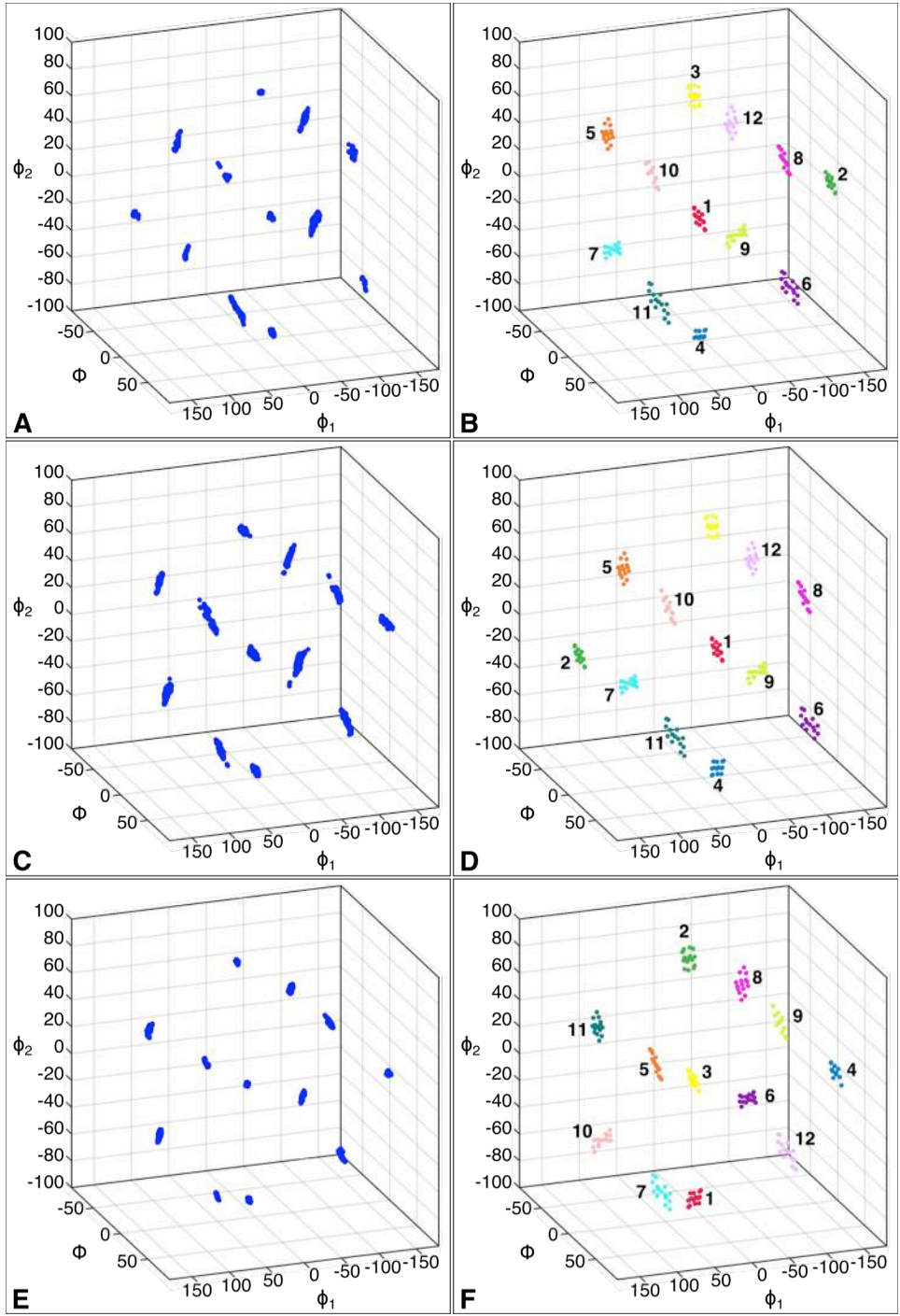

**Fig. A.1.** The experimental orientations are plotted by Euler angles ($\varphi_1$, $\Phi$, and $\varphi_2$) for sample 1 (A), and the calculated theoretical orientations are plotted with the variant orientation types separated by color and numbered (B). The experimental orientations are plotted by Euler angles ($\varphi_1$, $\Phi$, and $\varphi_2$) for sample 2 (D), and the calculated theoretical orientations are plotted with the variant orientation types separated by color and numbered (C). The experimental orientations are plotted by Euler angles ($\varphi_1$, $\Phi$, and $\varphi_2$) for sample 3 (e), and the calculated theoretical orientations are plotted with the variant orientation types separated by color and numbered (f).

The NiTi B2 lattice parameter of 3.015 Å was used (Otsuka et al., 1971). For each B19′ orientation, the minimum misorientations between the predicted B19′ orientation and all experimental B19′ orientations were calculated (i.e., the misorientation between the predicted B19′ orientation and its nearest experimental neighbor). These minimum misorientation angles were summed for all predicted B19′ orientations. The B2 orientation that produced the smallest summed minimum misorientation angle was accepted as the average B2 orientation for that sample.

Once the average B2 orientation was known for each sample, the experimental orientations could be assigned to a variant. In **Fig. A.1A,C,E**, the experimental orientations for samples 1, 2, and 3, respectively, are plotted by the Euler angles ($\varphi_1$, $\Phi$, and $\varphi_1$) that define each orientation. In **Fig. A.1B,D,F**, the theoretical B19′ twin orientations based on the calculated average B2 orientation are shown with the variant types assigned to each cluster. The theoretical variant orientation clusters are colored differently for clarity. By comparing the experimental and theoretical clusters for each sample, we were able to assign a variant type to each experimental orientation cluster. The volume of each variant was calculated as the sum of the volumes of all of the orientations assigned to that variant.

### A.3.3 Misorientation-strain profiles

The lattice misorientation magnitudes (**Figs. 9 11**, and **C.5**) and misorientation axes (**Fig. 13**, **Table 2**, and **Table C.1**) were calculated for the favorable variant in samples 1 and 2 by following the ff-HEDM least squares analysis procedure as presented in (Pagan and Miller, 2016). This procedure uses **Equation A.3** to calculate the lattice misorientation magnitude $\Delta\varphi$ and misorientation axis $\boldsymbol{t}$ from the nominal or initial lattice plane normal $\bar{\boldsymbol{n}}$ and the extreme lattice plane normals of the rotated lattice $\boldsymbol{n}_S$ and $\boldsymbol{n}_E$, where $S$ and $E$ refer to start and end, respectively.

$$\boldsymbol{v} = \boldsymbol{n}_E - \boldsymbol{n}_S = (\varphi \boldsymbol{t} \times \boldsymbol{I}) \cdot \bar{\boldsymbol{n}} \tag{A.3}$$

The vector $\boldsymbol{v}$ is the vector of reflection extension or "spreading." The lattice plane normals $\bar{\boldsymbol{n}}$, $\boldsymbol{n}_S$, and $\boldsymbol{n}_E$ were calculated from the centroid, start, and end of each reflection, respectively, at each load step using the calibrated detector parameters and the initial, unloaded B19′ lattice parameters.

When performed on a single reflection, this calculation is not sufficient to uniquely identify the misorientation axis and magnitude—the analysis using a single reflection can only be used to identify the plane in which the misorientation axis lies. Therefore, a least squares minimization using at least two (*hkl*) types must be performed to uniquely identify the misorientation axis and magnitude (Pagan and Miller, 2016). This procedure requires that the start and end of each reflection are easily identifiable since they are inputs to the calculations. For this reason, only certain reflections corresponding to certain grains were used for the least squares minimization. Reflections that were misoriented or located in such a way that the individual endpoints of the reflections were masked were omitted from the analysis. The favorable variant inside three different grains in sample 1 was analyzed, and the favorable variant inside three different grains in sample 2 was analyzed. For each grain analyzed, the reflections from the {100}, {011}, {$\bar{1}$21}, and {$\bar{1}\bar{2}$1} rings were used, and the misorientation axis and magnitude were calculated using a least squares minimization. All other rings were of insufficient intensity, were not clearly misoriented (i.e., the plane normals were aligned with the lattice misorientation axis), overlapped into other Debye–Scherrer rings, or contained reflections that spread into the reflections of other grains. The reflections in sample 3 were not oriented in such a way where these requirements were met, so this analysis was not applied to sample 3 data.

The above procedure was used to identify the misorientation axis for each grain. The following procedure discusses how the strain and misorientation along the misorientation axis was analyzed for each crystal. **Fig. A.2A** shows all (100) reflections from the illuminated microstructure sample 1 at load step 4 plotted on an HEDM pole figure (Pagan and Miller, 2016). A box is drawn around the (100) reflections that correspond to the favorable variant orientations, and these reflections are shown in isolation in **Fig. A.2B**. Each reflection corresponds to the favorable variant in a specific grain. The reflection extension vector $\boldsymbol{v}$ (i.e., the vector along which the reflection is spread in $\omega, \eta$ space) and the start and end of the reflection, $n_S$ and $n_E$, are labeled. The reflection corresponding to one particular grain is segregated into "boxes" equally spaced in 1° increments along $\boldsymbol{v}$. The boxes represent bounds in $\omega$ and $\eta$. These $\omega, \eta$ bounds are applied to the raw ff-HEDM reflection in (**Fig. A.2C**), where lower and upper bounds in 2  now form the third dimension of the boxes, resulting in a full discretization of the reflection along the $\boldsymbol{v}$ (**Fig. A.2D**). The minimum and maximum $2\theta$ is just below and just above that of the (*hkl*) ring to which the reflection belongs. For each discretized portion of the reflection, the misorientation and the elastic lattice strain are calculated. The misorientation, $\varphi \boldsymbol{t}$, is measured relative to the overall reflection centroid ($\overline{\boldsymbol{n}}$) using **Equation A.3**. The elastic lattice strain, $\varepsilon^{(hkl)}$, reported in (**Figs. 9 11**, and **C.5**) was calculated using the conventional *d*-spacing analysis shown in **Equation A.4**.

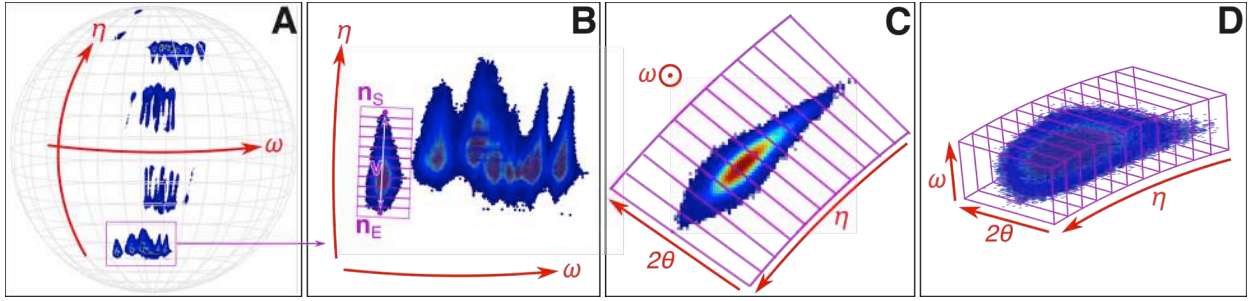

**Fig. A.2.** Explanation for the lattice strain versus misorientation calculations. (A) shows all (100) reflections from sample 1 at load step 4 plotted on an HEDM pole figure. The reflections corresponding to the favorable variant orientations are boxed in (A) and shown in close-up in (B), with one particular reflection segregated into $\omega, \eta$ boxes. These $\omega, \eta$ bounds are applied to the raw ff-HEDM reflection in (C), where the third dimension of the boxes is now formed by lower and upper bounds in $2\theta$, resulting in a full discretization of the reflection along the extension vector (D).

$$\varepsilon_{hkl} = \frac{d-d_0}{d_0}, \tag{A.4}$$

The initial *d*-spacing, $d_0$, is the average *d*-spacing of the reflection at zero load. This procedure allows one to calculate the lattice strain associated with a particular portion of a misoriented, or rotated, lattice.

The relative lattice misorientation was calculated using the initial, unloaded B19′ lattice parameters shown in **Table A.3**, calculated using the GSAS-II peak fitting module on the unloaded ff-HEDM data sets (Toby and Von Dreele, 2013).

**Table A.3.** Mean (fitting standard deviation) B19′ monoclinic lattice parameters.

| a | b | c | β |
|---|---|---|---|
| 2.8650(8) Å | 4.1221(3) Å | 4.671(2) Å | 97.217(9)° |

# Appendix B. Near-field high-energy diffraction microscopy analyses and visualizations

## B.1 Data collection

The same monochromatic X-ray energies set to the middle of the $K_\alpha$ absorption edges of holmium (55.618 keV) for samples 1 and 2 and ytterbium (61.332 keV) for sample 3 were used for nf-HEDM. The load steps for nf-HEDM data collection are indicated in **Table 1**. For nf-HEDM measurements, the X-ray beam was masked to 0.1 mm tall × 2 mm wide using slits. For sample 1, a 1×1×1 mm$^3$ volume was measured at the vertical center of the gage section by acquiring ten contiguous nf-HEDM measurements, vertically translating the sample in 0.1 mm intervals for each measurement. For samples 2 and 3, a 0.5×1×1 mm$^3$ volume was measured by acquiring five contiguous nf-HEDM measurements, vertically translating the sample in 0.1 mm intervals for each measurement. A Retiga 4000DC camera with 2048×2048 pixels and a 1.48×1.48 μm$^2$ pixel size was used to collect images of diffraction patterns on a LuAG:Ce scintillator. The near-field detector was characterized using an assembly of two offset 25×25×50 μm$^3$ gold crystals, where each crystal contained several crystal orientations, and the calibration parameters were optimized using HEXRD (Bernier, 2017). The distance from the sample to the near-field detector was 7.28 mm for samples 1 and 2 and 7.86 mm for sample 3. Exposures were taken at each 0.5° increment of sample $\omega$ rotation over a full 360° rotation, resulting in 720 recorded detector images per 0.1 mm vertical layer per nf-HEDM data collection. A median background subtraction was used for each layer.

Microcomputed tomography (μCT) measurements of the gage sections were also recorded to provide spatial bounds to the nf-HEDM reconstructions. Specifically, because the sample will absorb some of the X-rays when illuminated, there will be intensity contrast between direct images taken with the sample in place and images taken without the sample in place (i.e., of the incident beam). As a result, this contrast can be used to identify the bounds of the sample and use them to confine the nf-HEDM reconstructions in space. μCT measurements were taken just before each nf-HEDM measurement using the exact same setup, detector, and X-ray beam energy, only removing the beam stop and taking 360 one-second exposures during the full sample rotation (1 exposure per degree of rotation). A "bright field" image (i.e., a one-second exposure with the X-ray beam on, but without the sample in the beam) was used to normalize the μCT data for background.

**B.2 Data analysis**

The nf-HEDM technique provides spatial and orientation information that can be used to make 3D reconstructions of the grain network. The nf-HEDM data analyses were also performed using the HEDM analysis software suite HEXRD (Bernier, 2017). This software follows the typical nf-HEDM analysis procedure of forward-modeling orientations onto a virtual detector (Poulsen, 2012, 2004; Schmidt, 2014). In the nf-HEDM forward-modeling procedure, a 3D voxel grid is constructed of the sample space, where the voxel dimensions are limited by the spatial resolution of the experimental setup. In our case, the spatial resolution of the grain maps was limited by our experimental setup to be ~2 μm. However, we found a voxel grid size of 5×5×5 μm$^3$ resulted in sufficiently detailed and less computationally expensive reconstructions. For a particular voxel, each orientation from a list of orientations is forward modeled, and the locations (in $\omega$, $\eta$, and $2\theta$ as defined in **Fig. 4**) of the simulated reflections are stored. The completeness of the orientation at that voxel is equal to the number of virtual reflections that matched experimental reflections divided by the number of virtual reflections (Gotz et al., 2007). The orientation with the highest completeness is assigned to that voxel. This procedure is repeated for each voxel.

The above procedure requires a list of orientations to forward model (i.e., to try). Because orientations can be directly inverted from ff-HEDM data sets (as opposed to require forward modeling), the typical convention is to first index a complementary ff-HEDM data set collected at the same load step and use the indexed orientations to forward model for the nf-HEDM reconstruction. Here, we used the orientations indexed from the ff-HEDM data sets but also included orientation distributions spanning ±5° in 1° intervals, where each distribution centered on an orientation indexed from the ff-HEDM data set. This is advised for reconstructions where small changes in orientation (e.g., intragranular misorientation) is of interest (e.g., (Oddershede et al., 2015; Winther et al., 2017)), because the near-field data is often more sensitive to spatial variations in orientation than the far-field detector.

To improve the accuracy of the reconstructions of the free surfaces of the samples, the μCT data were analyzed using the Inverse Radon Transform in the SciPy library (Jones et al., 2001), and the reconstructions were overlaid with the nf-HEDM reconstructions, and any material outside of the free surfaces measured by μCT was subtracted prior to final visualization.

**B.3 Data visualization**

In this study, we used large disparities in completeness (see **Section A.1** for definition of completeness) to separate fully detwinned regions from twinned regions of the gage sections. If the domains in the material are much smaller than the spatial resolution, then the maximum completeness at these locations is significantly lower than the average completeness values for the rest of the reconstruction. This completeness disparity is demonstrated in **Fig. B.1**. **Fig. B.1B** shows a 2D slice of the 3D reconstruction for sample 3 (**Fig. 8**) with orientations colored according to the IPF colormap shown in **Fig. B.1A**. At the top of the reconstruction, there are large, fully detwinned grains consisting of only the favorable variant, and at the bottom of the reconstruction, it is clear that the microstructure is still twinned. In the twinned region, the reconstruction is quite poor, because the twin domains are below the spatial resolution of the technique. **Fig. B.1C** shows the corresponding completeness, map for this reconstruction. At the top of the reconstruction, the fully detwinned microstructure has high completeness values of 60%–70%. At the grain boundaries, the completeness is slightly lower at 40–50%. At the bottom of the reconstruction, the twinned microstructure has very low completeness values of 20–40%. **Fig. B.1D** shows this same reconstruction but with a completeness cutoff threshold of 40% used to remove low completeness data, which isolates the fully reoriented region from the still-twinned region. Applying this analysis over the entire reconstructed volume results in isolation of the fully reoriented region from the twinned region. This analysis methodology of using completeness thresholding to isolate different regions of microstructure can in theory be applied to other situations as well; microstructures that have (or have not) undergone severe plastic deformation, grain refinement, regions with large void or microcrack concentrations, etc. will exhibit analogous completeness disparities in nf-HEDM data.

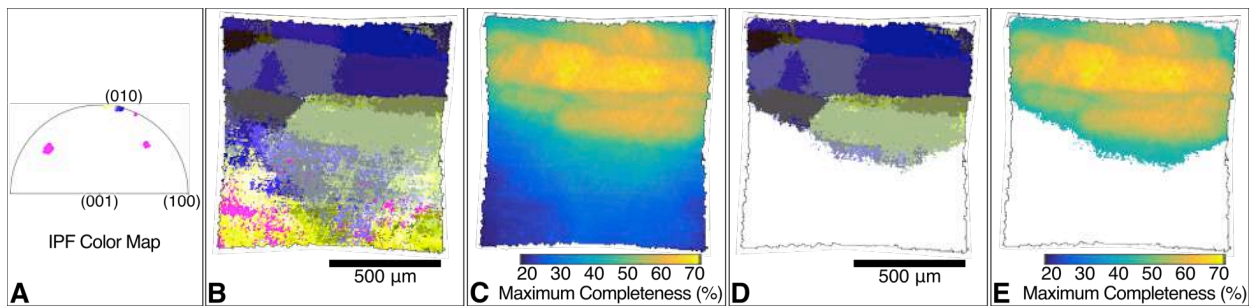

**Fig. B.1.** A 2D slice from the 3D reconstruction for sample 3 shown in **Fig. 17** is shown in shown in (B) where the colors indicate orientation according to the IPF colormap in (A). (C) is the corresponding completeness map. The same reconstruction slice (D) and corresponding completeness map (E) is shown after a completeness threshold of 40% is applied.

All 2D grain reconstruction visualizations were generated using MTEX (Bachmann et al., 2011), and the 3D using ORS Dragonfly 2.0 (*Dragonfly 2.0*, 2016) (**Figs. 8b, 9b, 13b, 17b, and B1**).

**Appendix C. Supplementary results for samples 2 and 3**
**C.1 Sample 2**

Selected frames from DIC analysis for sample 2 in **Fig. C.1** show the 2D surface strains of one of the four faces of the $1\times1\times8$ mm$^3$ gage section using a uniform 11% maximum strain scale bar (**Fig. C.1A**) and narrower frame-specific scale bars to contrast local heterogeneities (**Fig. C.1B**). The white boxes in **Fig. C.1** outline the portion of the gage section that was illuminated by X-rays during ff-HEDM measurements at load steps 0–5. The DIC frames corresponding to ff-HEDM measurements are labeled by the load step numbers at the top of each frame, correlating with **Fig. 2B**.

**Fig. C.2A** shows the evolution of the monoclinic texture, where the texture is shown in equal area projection IPFs with respect to the loading axis and the colors correspond to relative volume. The schematic at the top of **Fig. C.2A** indicates how the orientations of each variant are "clustered" in orientation space. The relative volume fraction of each of the 12 variants were binned at each load step and are shown in the histograms in **Fig. C.2B**, where the arrows indicate whether the volume of each variant increased or decreased from the previous load step. At the start of region III (green background), only three variants remain in the microstructure: variants 4, 5, and 12. Following the twin pair definitions in (Bhattacharya, 2003; Hane and Shield, 1999), this combination of variants corresponds to Mode D Type I/II twins consisting of variants 4 and 5, 4 and 12, and/or 5 and 12. A closer examination of the reflection symmetry of the variant clusters showed that these twins are ($\bar{1}11$) Type I twins composed of variants 4 and 12, and [$\bar{2}11$] Type II twins composed of variants 4 and 5. At the end of the plateau, the microstructure has fully reoriented to the favorable variant 4.

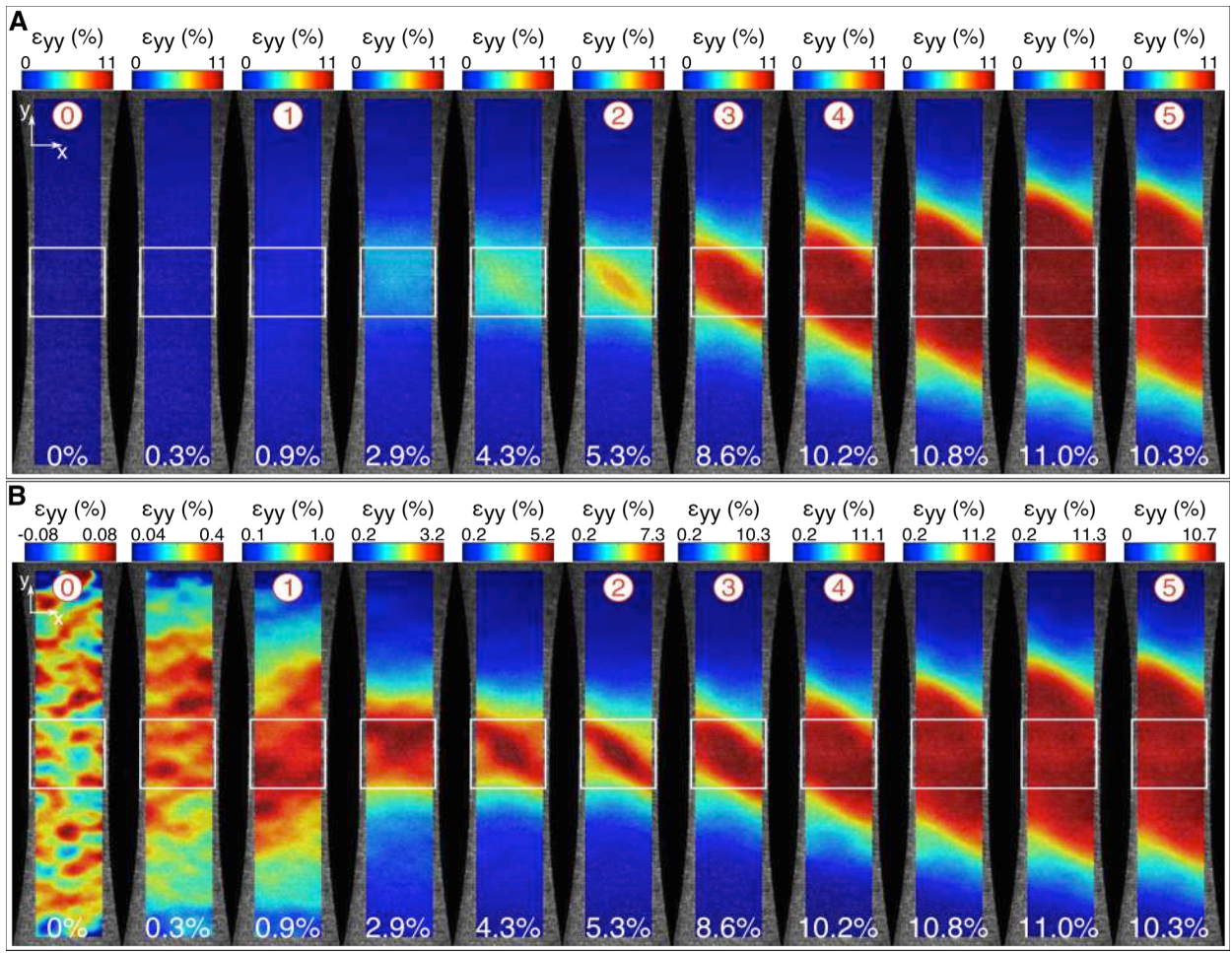

**Fig. C.1.** Selected frames from the DIC strain analysis for sample 2. The strain is colored using the scale bars indicated above each frame, with the frames shown in (A) using a uniform scale of 0%–11% strain and the frames in (B) using varying scales to contrast strain variations. The white boxes indicate the region that was illuminated by X-rays during ff-HEDM measurements, and the mean strain in the loading direction in the illuminated region is given at the bottom of each frame. The circled numbers indicate the frames at load steps 0–5.

The nf-HEDM reconstruction for sample 2 reveals the internal grain network as well as some intragranular deformation (**Fig. C.3B**). The 1×1×0.5 mm$^3$ illuminated volume during the nf-HEDM measurement is indicated by the boxed region in **Fig. C.3A**. The 3D reconstruction of the illuminated volume in **Fig. C.3B** is a spatially resolved orientation map of the fully reoriented favorable variant. Two 2D slices of the 3D reconstruction are also shown in **Fig. C.3B**. The colors of the orientations correspond to (*hkl*) as shown in the "Reconstructed Orientations" IPF in **Fig. C.3C**. A comparison of the "Orientations Present" (colored by relative volume) and the "Reconstructed Orientations" (colored by (*hkl*)) in **Fig. C.3C** show that all of the orientations

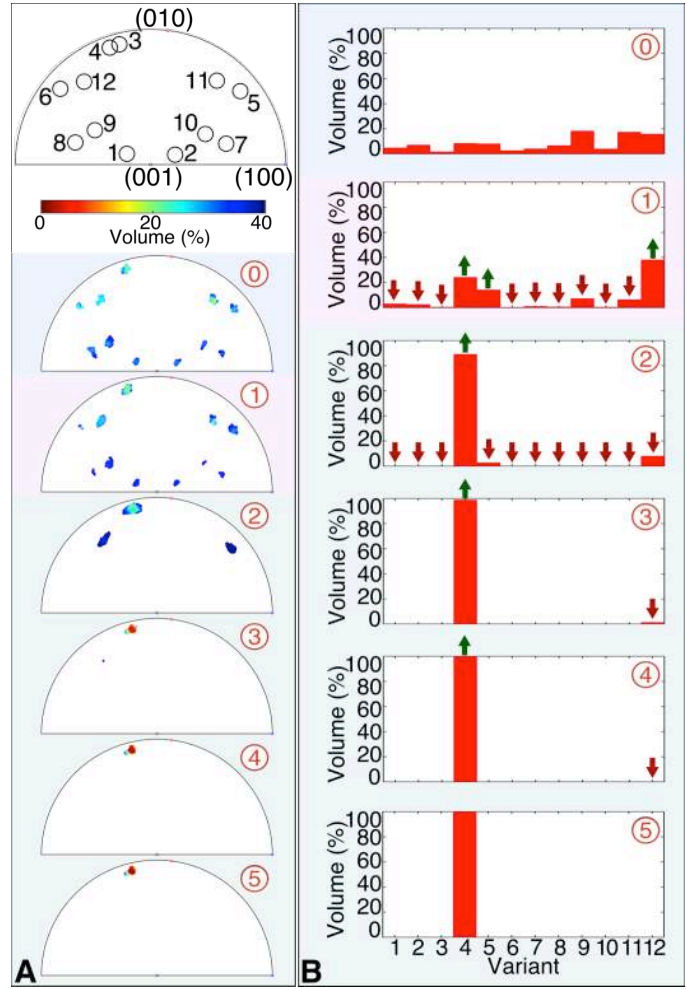

**Fig. C.2.** Texture evolution of sample 2. (A) The texture is shown at each load step colored by volume. (C) The volume of each variant at each load step is shown in histograms, with arrows indicating whether the volumes increased or decreased from the previous load step.

present in the microstructure have been reconstructed in **Fig. C.3B**.

The ff-HEDM data sets were used to make subgrain-resolved measurements of lattice rotation vs. corresponding elastic strain experienced by the favorable variant inside three different grains. The raw ff-HEDM data corresponding the four Bragg reflection types, (100), ($10\bar{1}$), ($\bar{1}\bar{2}1$), and ($\bar{1}21$), from the favorable variant are shown together at load steps 0–5 in **Fig. C.4**, and the resultant elastic lattice strain vs. misorientation profiles are shown in **Fig. C.5**. Notice that the ($\bar{1}\bar{2}1$) and ($\bar{1}21$) measurements from grains 1 and 3 are missing at load steps 0–1 in **Fig. C.5**. The ($\bar{1}\bar{2}1$) and ($\bar{1}21$) Bragg reflections did not exist above the detector noise level for these two grains until load step 2 (see **Fig. C.4**). The misorientation magnitudes and axes are provided in **Table C.1**.

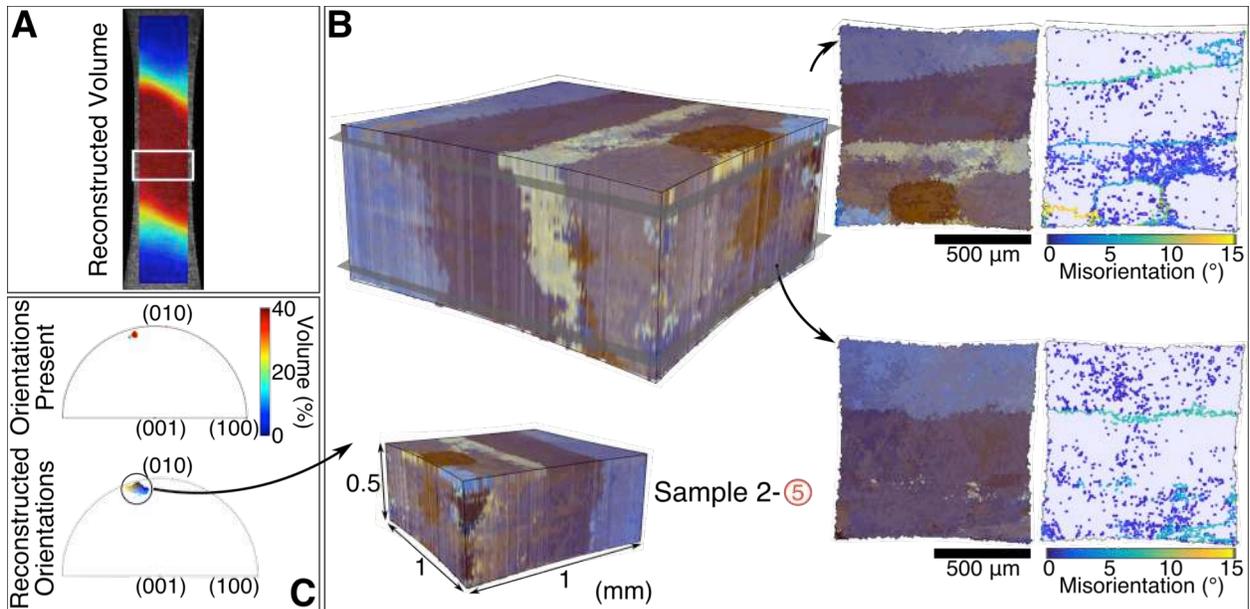

**Fig. C.3.** 3D, in situ reconstruction of the detwinned microstructure inside the LDB for sample 2 at load step 5. The 1×1×0.5 mm$^3$ illuminated volume during the measurement is indicated by the boxed region in (A). The 3D orientation map of the fully reoriented favorable variant is shown in (B), where the colors indicate orientation according to the "Reconstructed Orientations" IPF colormap shown in (C). In (C), the "Orientations Present" IPF shows all of the orientations present in the illuminated volume, colored by relative volume, and the "Reconstructed Orientations" IPF shows all of the orientations reconstructed in (B) colored by ($hkl$).

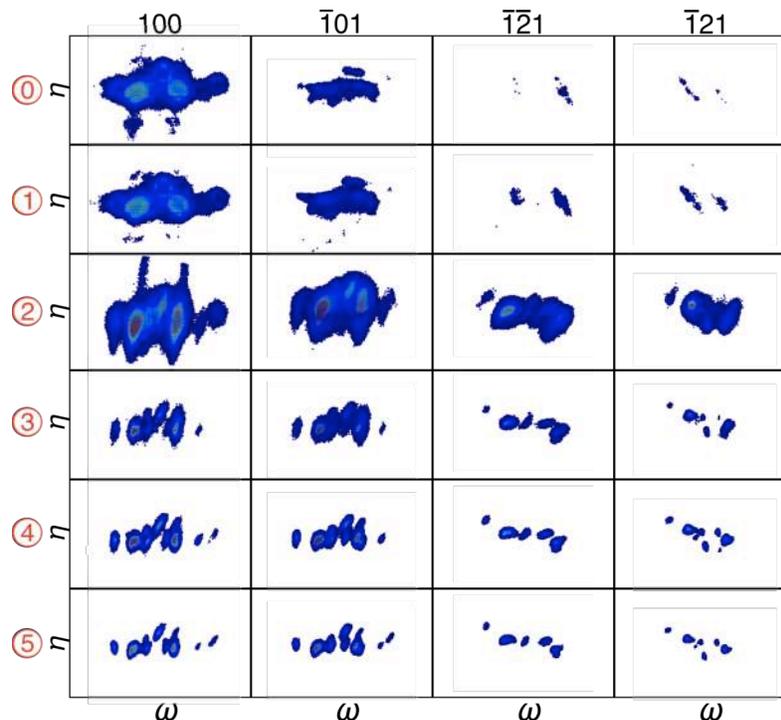

**Fig. C.4.** (100), (10$\bar{1}$), ($\bar{1}\bar{2}$1), and ($\bar{1}$21) Bragg reflections from the favorable variant in sample 2 at load steps 0–5.

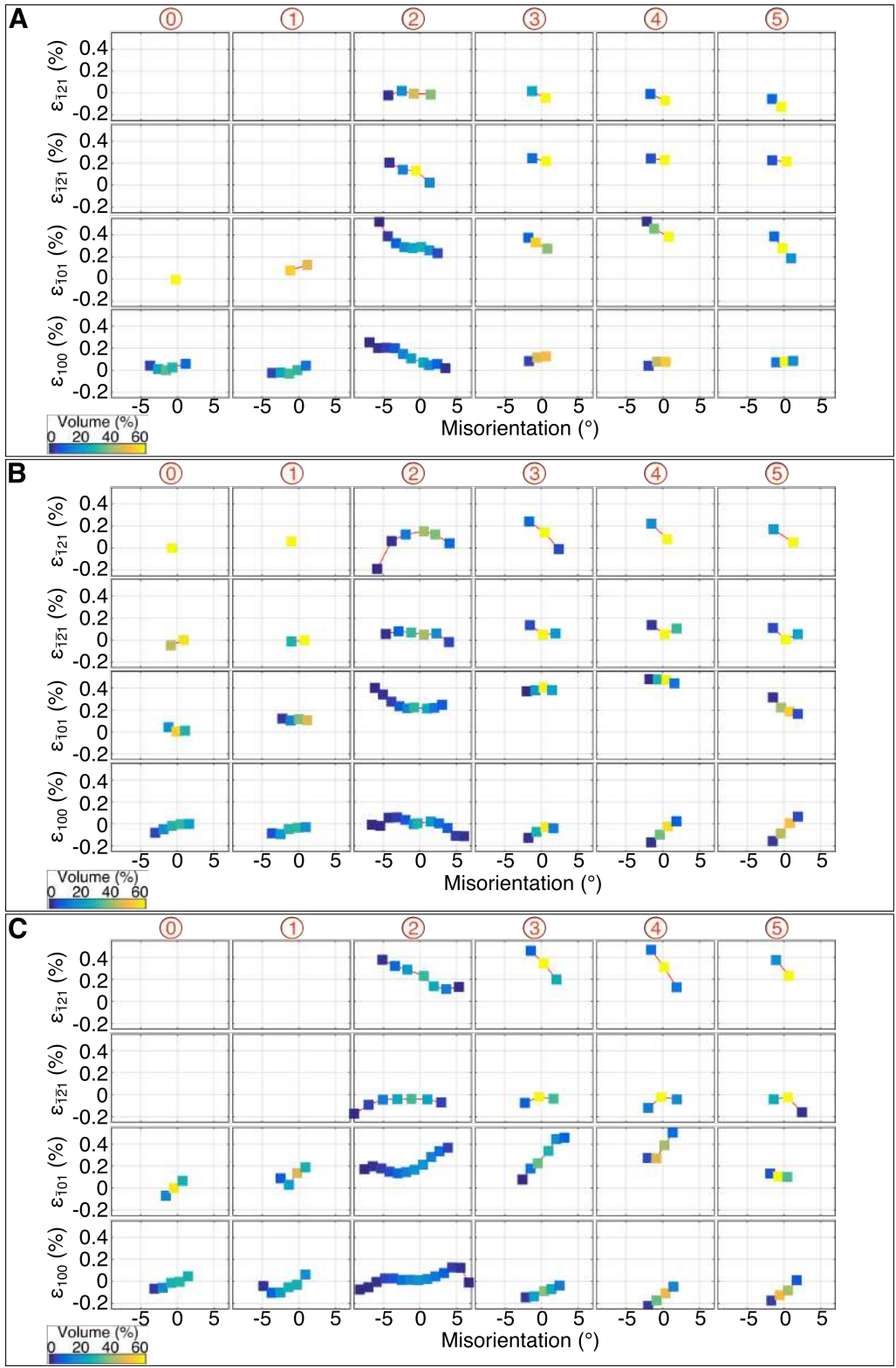

**Fig. C.5.** Elastic lattice strain vs. misorientation experienced by the favorable variant within grain 1 (A), grain 2 (B), and grain 3 (C) in sample 2 calculated from the (100), (10$\bar{1}$), ($\bar{1}\bar{2}$1), and ($\bar{1}$21) Bragg reflections.

**Table C.1.** Misorientation magnitudes and axes for each grain included in the misorientation-lattice strain analysis for sample 1. The axes ($t$) are reported in the laboratory coordinate system as shown in **Fig. 4,** as are the ($hkl$).

|  | Misorientation, $\varphi$ (°) | Misorientation Axis, $t$ | Misorientation Axis ($hkl$) |
|---|---|---|---|
| Grain 1 | 10.42 | [0.16 0.08 0.98] | ($\overline{0.08}$ 1 $\overline{0.03}$) |
| Grain 2 | 10.99 | [0.05 0.13 0.99] | (0.03 1 $\overline{0.03}$) |
| Grain 3 | 12.41 | [$\overline{0.03}$ 0.08 1.00] | ($\overline{0.01}$ 1 0.04) |

## C.2 Sample 3

Selected frames from DIC analysis in **Fig. C.6** for sample 3 show the 2D surface strains of one of the four faces of the 1×1×1 mm$^3$ gage section using a uniform 12% maximum strain scale bar (**Fig. C.6A**) and narrower frame-specific scale bars to contrast local heterogeneities (**Fig. C.6B**). The white boxes in **Fig. C.6** outline the portion of the gage section that was illuminated by X-rays during ff-HEDM measurements at load steps 0–4. The DIC frames corresponding to ff-HEDM measurements are labeled by the load step numbers at the top of each frame, correlating with **Fig. 2C**. Due to the different specimen geometry of sample 3 than that of samples 1 and 2, the LDB interfaces remained within the illuminated region of the gage section where the load was concentrated.

**Fig. C.7A** shows the evolution of the monoclinic texture, where the texture is shown in equal area projection IPFs with respect to the loading axis and the colors correspond to relative volume. The schematic at the top of **Fig. C.7A** indicates how the orientations of each variant are "clustered" in orientation space. The relative volume fraction of each of the 12 variants were binned at each load step and are shown in the histograms in **Fig. C.7B**, where the arrows indicate whether the volume of each variant increased or decreased from the previous load step. The behavior of the variant volumes are similar to that of samples 1 and 2, with the favorable variant plus two lower volume variants becoming dominant before fully detwinning to the favorable variant. However, the changes of the variant volumes occur later with respect to the macroscopic loading for sample 3 compared to samples 1 and 2. For example, the microstructure did not reduce to three variant types until load step 3, somewhat far into the stress plateau, whereas the microstructure reduced to three variant types at the beginning of the plateaus for sample 1 (load step 3 in **Fig. 6B**) and sample 2 (load step 2 in **Fig. C.2B**). The difference comes from the fact that the LDB is "narrower" for sample 3 because of the shorter gage section specimen geometry,

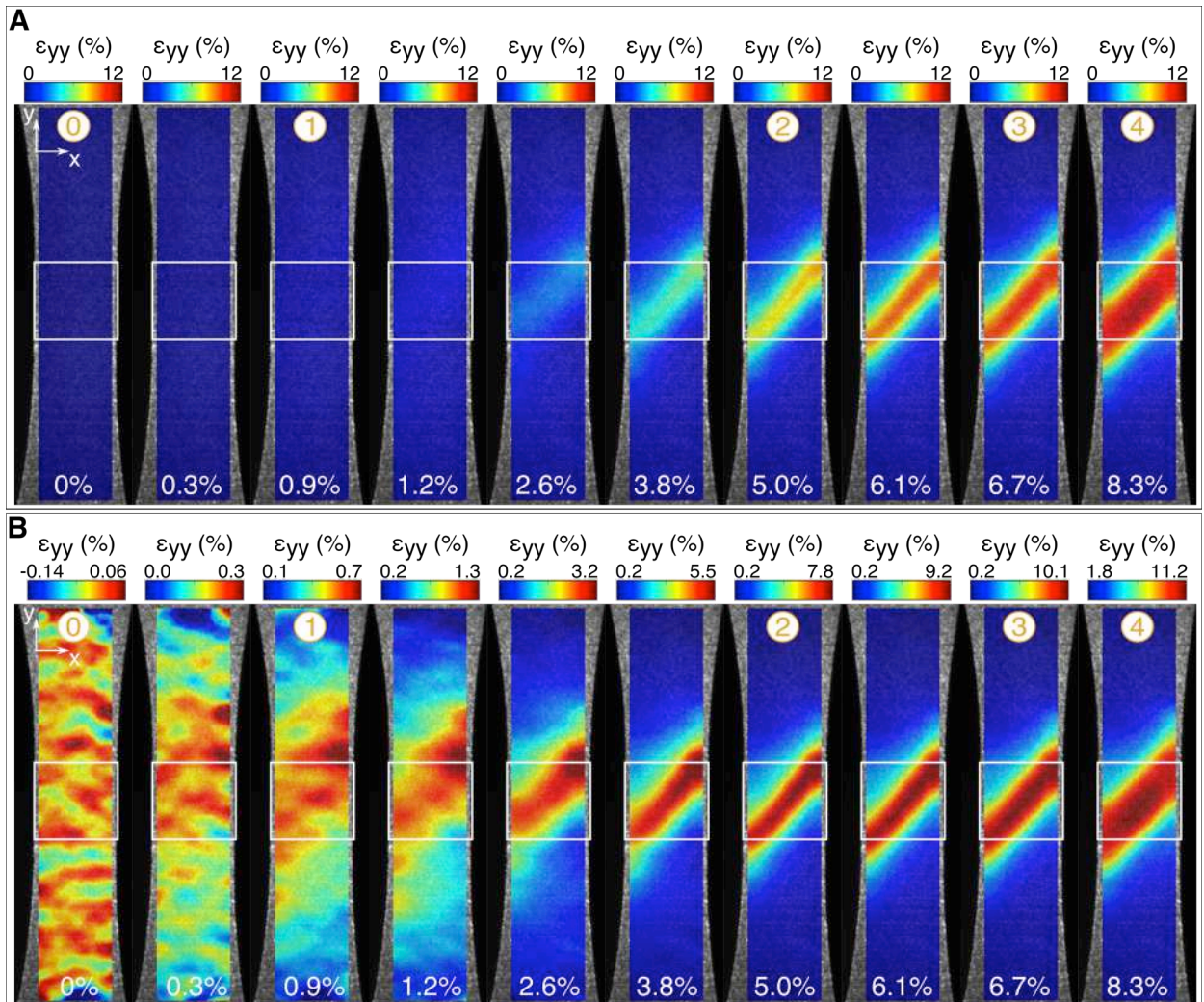

**Fig. C.6.** Selected frames from the DIC strain analysis for sample 3. The strain is colored using the scale bars indicated above each frame, with the frames shown in (A) using a uniform scale of 0%–12% strain and the frames in (B) using varying scales to contrast strain variations. The white boxes indicate the region that was illuminated by X-rays during ff-HEDM measurements, and the mean strain in the loading direction in the illuminated region is given at the bottom of each frame. The circled numbers indicate the frames at load steps 0–4.

meaning that regions outside of the LDB remained within the illuminated volume for much longer (see. **Fig. C.6**). At the end of the plateau, only three variants remain in the microstructure: variants 2, 6, and 10. Following the twin pair definitions in (Bhattacharya, 2003; Hane and Shield, 1999), this combination of variants corresponds to Mode D Type I/II twins consisting of variants 2 and 6, 2 and 10, and/or 6 and 10. A closer examination of the reflection symmetry of the variant clusters showed that these twins are $(\bar{1}11)$ Type I twins composed of variants 2 and 6, and $[\bar{2}11]$ Type II twins composed of variants 2 and 10. In the case of sample 3, the favorable variant is variant 2. Again, for the grain orientations (in the B2 austenite sense) present in sample 3, variant 2 is the variant that maximizes the deformation (i.e., work output) in the loading direction. Because all three samples contained different grain orientations, all three samples preferred different favorable variants.

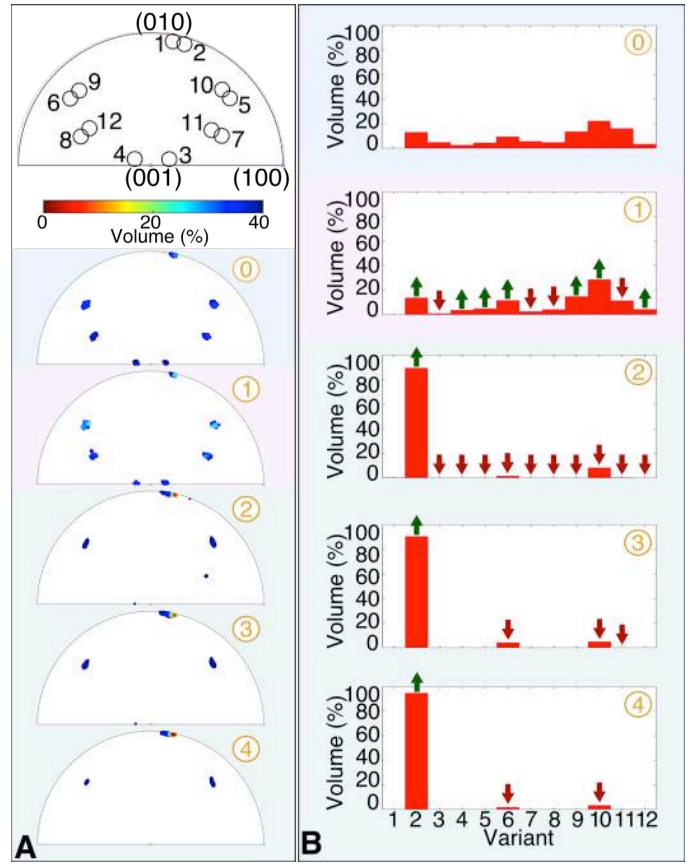

**Fig. C.7.** Texture evolution of sample 3. (A) The texture is shown at each load step colored by volume. (C) The volume of each variant at each load step is shown in histograms, with arrows indicating whether the volumes increased or decreased from the previous load step.